\documentclass[a4paper, 11pt]{amsart}

\usepackage{amsmath, amssymb, amsthm}
\usepackage[english]{babel}
\usepackage[T1]{fontenc}
\usepackage{url}
\usepackage{graphicx} 
\newcommand{\R}{\mathbb R}

\def\be#1\ee{\begin{equation}#1\end{equation}}
\newcommand{\fer}[1]{(\ref{#1})}
\newtheorem{theorem}{Theorem}[section]

\newtheorem{remark}[theorem]{Remark}

\newcommand{\bq}{\begin{equation}}
\newcommand{\eq}{\end{equation}}
\newenvironment{equations}{\equation\aligned}{\endaligned\endequation}
\def\bqa{\begin{eqnarray}}
\def\eqa{\end{eqnarray}}

\def\e{\epsilon}

\def\t{\tau}

\newcommand{\bd}{\begin{displaymath}}
\newcommand{\ed}{\end{displaymath}}
\newcommand{\ba}{\begin{eqnarray}}
\newcommand{\ea}{\end{eqnarray}}

\def\R{\mathbb{R}}

\theoremstyle{plain}

\title{Human Behavior And Lognormal Distribution. A Kinetic Description}

\author{Stefano Gualandi and Giuseppe Toscani}

\thanks{Department of Mathematics  of the University of Pavia and IMATI CNR, Italy.  e.mail: stefano.gualandi@unipv.it, giuseppe.toscani@unipv.it. 
}

\date{\today}

\begin{document}
\maketitle

\begin{center}\small
\parbox{0.85\textwidth}{

\textbf{Abstract.} In recent years it has been increasing evidence that lognormal distributions are widespread in  physical and biological sciences, as well as in various phenomena of  economics and social sciences. In social sciences, the appearance of lognormal distribution has been noticed, among others, when looking at body weight,
and at women's age at first marriage.
 Likewise, in economics, lognormal distribution appears when looking at consumption in a western society,
 at call-center service times,
 and others.
The common feature of these situations, which describe the distribution of a certain people's hallmark, is the presence of a desired target to be reached by repeated choices.
In this paper we discuss a possible explanation of lognormal distribution forming in human activities by resorting to classical methods of statistical mechanics of multi-agent systems.
 The microscopic variation of the hallmark around its target value, leading to a linear Fokker--Planck type equation with lognormal equilibrium density, is  built up introducing as main criterion for decision  a suitable value function in the spirit of the prospect theory of Kahneman and Twersky.
\medskip

}
\end{center}

\section{Introduction}

The study of random variations that occur in the data from many scientific disciplines often show 
more or less skewed probability distributions. Skewed distributions are particularly common
when the measured values are positive, as it happens, for example, with species abundance \cite{Hir,Lop}, lengths of latent periods of infectious diseases \cite{Kon,Sar1,Sar2}, and distribution of mineral resources in the Earth's crust \cite{Ahr,Mal,Raz}. Skewed distributions often closely fit the lognormal distribution \cite{Aic,Cro,Lim}. The list of phenomena which fit lognormal distribution in natural sciences is quite long, and the interested reader can have an almost complete picture about them by reading the exhaustive review paper by Limpert, Stahel and Abbt \cite{Lim}. 

In addition to samples from physical and biological sciences, a relevant number of phenomena involving measurable quantities of a population and fitting lognormal distribution comes from social sciences and economics, areas where it can be reasonably assumed that the appearance of this distribution is a consequence of a certain human behavior.  

Among others, a good fitting has been observed while looking at the distribution of body weight \cite{BC}, at women's age at first marriage \cite{Pre}, at drivers behavior \cite{JJ}, or, from the economic world, when looking at consumption in a western society \cite{BBL}, at the size of cities \cite{BRS}, and at call-center service times \cite{Brown}. 

Most of the scientific work in this area of research is mainly devoted to understand at best the possible reasons which justify the underlying lognormal distribution, and to estimate its parameters, while dynamical mathematical models trying to explain the formation of lognormal distribution are usually not dealt with. 

In this paper we will try to close this gap by showing that the aforementioned phenomena in social sciences and economics can be reasonably well described in terms of a partial differential equation of Fokker--Planck type for the density $f = f(w,t)$ of agents which have the (positive) value of the hallmark under consideration equal to $w$ at time $t\ge 0$.  This Fokker--Planck equation, which will be the main object to study, takes the form

\begin{equation}\label{FPori}
 \frac{\partial f(w,t)}{\partial t} = \frac \lambda 2 \frac{\partial^2 }{\partial w^2}
 \left(w^2 f(w,t)\right )+ \frac \gamma 2
 \frac{\partial}{\partial w}\left(  w\, \log \frac w{\bar w_L} f(w,t)\right).
 \end{equation}
In \fer{FPori} $\lambda, \gamma$ and $\bar w_L$ are positive constants  closely related to the typical quantities of the phenomenon under study.  In view of the fact that the independent variable $w$ is non-negative, the initial value problem for the Fokker--Planck equation \fer{FPori} is usually coupled with suitable boundary conditions at the point $w=0$ \cite{PT13,FPTT}. The equilibrium density of the Fokker--Planck equation \fer{FPori} is given by the lognormal density
 \be\label{equili}
f_\infty(w) = \frac 1{\sqrt{2\pi \sigma}\, w} 
\exp\left\{ - \frac{(\log w - \kappa)^2}{2 \sigma}\right\},
 \ee 
where 
 \be\label{pa}
 \sigma = \frac \lambda\gamma,  \quad \kappa = \log \bar w_L - \sigma.
 \ee
Moreover, for any given initial distribution $f_0(w)$ of agents, convergence to the lognormal equilibrium is shown to hold exponentially fast in time with explicit rate (cf. Section \ref{trend}). 

The Fokker--Planck equation \fer{FPori} has been first derived in the economic context in  \cite{GT17}.
There, this equation has been obtained starting from a detailed explanation of the possible motivations behind the forming of a lognormal (steady) distribution in agents service times of call centers (previously observed by Brown and others \cite{Brown}), and resorting to the well-consolidated methods of statistical mechanics \cite{BKS,CFL,NPT,PT13}. 

The approach used in  \cite{GT17} has its roots in kinetic theory. This approach is robust in particular when modeling socio-economic problems.  Indeed, mathematical modeling of social and economical phenomena in multi-agent systems became in the last twenty years a challenging and productive field of research involving both applied mathematicians and physicists. In economics, the formation of Pareto curves in wealth distribution of western countries has been one of the main issues studied in various aspects \cite{ChaCha00,CCM,ChChSt05,CoPaTo05,DY00,GSV,SGD}. Likewise, in social sciences, the investigation of statistical mechanics of opinion formation played a leading rule \cite{BN2,BN3,BN1,BeDe,Bou,Bou1,Bou2,CDT,DMPW,GGS,GM,Gal,GZ,SW,To1}. 

Connections of kinetic modeling of social and economical phenomena with classical kinetic kinetic theory of rarefied gases are well-established. A recent discussion can be found in  \cite{GT-ec}. There, both the appearance of the classical Fokker--Planck equation and of its equilibrium, given by a normal density, are justified by analyzing the Kac caricature of Maxwell molecules \cite{Kac59}.

 Among others, previous research in the field of statistical mechanics, which revealed unexpected similarities with the problem of service times duration in call-centers \cite{GT17}  was  the statistical description of agents acting in a simple financial market, by taking into account specific behavioral aspects. A kinetic approach to this leading problem in economics has been proposed by Pareschi and Maldarella in  \cite{MD}. There the authors, to investigate the price formation of a good in a multi-agent market, introduced a kinetic model for a multi-agent system consisting of  two different trader populations, playing different rules of trading, and possibly changing their point of view. The kinetic description was inspired by the microscopic Lux--Marchesi model \cite{LMa,LMb} (cf. also  \cite{LLS,LLSb}). 
 
 The  connection with the present problem is mainly related to the trading rules, which were assumed to depend on the opinion of traders through a kinetic model of opinion formation recently introduced in  \cite{To1}. Also, psychological and behavioral components of the agents, like the way they interact with each other and perceive risks, which may produce non-rational behaviors, were taken into account. This has been done by resorting, in agreement with the prospect theory by Kahneman and Twersky \cite{KT,KT1}  to interactions containing a suitable \emph{value function}. 

The analysis of  \cite{MD} enlightens the importance of the human behavior pioneered in  \cite{Zipf}  (cf. \cite{BHT,BCKS}), which is there reproduced by the nonlinear value function, in the microscopic mechanism leading to the underlying kinetic equation, able to reproduce the macroscopic evolution of the market. Also, \emph{mutatis mutandis}, it suggests how the presence of a suitable value function can justify at a microscopic level the mechanism of formation of the service time distribution \cite{GT17}. 

The leading idea in  \cite{GT17}  can be expressed by this general principle. For a certain specific hallmark of the population of agents, measured  in terms of a positive value $w\in \R_+$, agents have the objective to reach a target value, corresponding to a certain fixed value $\bar w$.  This value could be reached by repeated upgrading, which correspond to microscopic interactions. However, the upgrade of the actual value towards the target value is different depending of the actual state of the agents. If the value $w$  is less than the target value $\bar w$,  to get closer to it is much more satisfactory than in the opposite situation. 

To clarify the meaning of the previous assertion, let us take into account the case in which the target value $\bar w$ to be reached is the departure time of a train, that often has some delay. Then, given a time interval of five minutes, the mood of the traveller  who is going to the train station, will be completely different once he will realize that the station will be reached five minutes before or after the departure time.
For example, by referring to the problem of the characterization of a call center service time, treated in  \cite{GT17},  an agent is more relaxed if he made a service  staying below the expected mean time fixed by the firm to close the operation, than in the opposite situation.  

In the forthcoming Section \ref{model}, we shall introduce a linear kinetic model for a multi-agent system, in which agents can be characterized in terms of a certain hallmark that can be measured in terms of a non nonnegative quantity (the weight in grams, the age of first marriage in years, and so on), and subject to microscopic interactions which describe the microscopic rate of change of the value of the hallmark itself, according to the previous general principle. 
As we shall see, the relevant mechanism of the microscopic interaction is given by resorting to a suitable value function, in the spirit of the analysis of Kahneman and Twersky\cite{KT,KT1}, which reproduces at best the human behavior.
 
 Then in Section \ref{quasi} we will show that in a suitable asymptotic
procedure (hereafter called \emph{quasi-invariant}  limit) the solution to the kinetic model tends towards the solution of the Fokker-Planck type equation \fer{FPori}.  

Similar asymptotic analysis was performed in   \cite{CPP,DMTb} for  a kinetic model for the distribution of wealth in a simple market economy subject to microscopic binary trades in presence of risk, showing formation of steady states with Pareto tails, in  \cite{TBD} on kinetic equations for price formation, and in  \cite{To1} in the context of opinion formation in presence of self-thinking. A general view about this asymptotic passage from  kinetic equations based on general interactions  towards Fokker--Planck type equations can be found in  \cite{FPTT}. Other relationships of this asymptotic procedure with the classical problem of the \emph{grazing collision limit} of the Boltzmann equation in kinetic theory of rarefied gases have been recently enlightened in  \cite{GT17}.

Once the  Fokker--Planck equation \fer{FPori} has been derived, the main examples quoted in this introduction will be collected in Section \ref{examples}, together with a detailed explanation  of the relevant mechanism which leads to the typical microscopic interaction in terms of the value function. 

In Section \ref{trend} we will further discuss various mathematical results concerned with the large-time behavior of the solution to this Fokker--Planck equation. In particular, we will show that convergence to equilibrium is exponentially fast in time, thus justifying \emph{a posteriori} the consistency of the model.

Last, resorting to some examples taken from real data sets, we will verify in Section \ref{numer}  that the kinetic approach provides an accurate description of these skewed human phenomena.

\section{The kinetic model}\label{model}

Let us consider a population of agents, which can be considered homogeneous with respect to some hallmark, that can be measured in terms of some standard positive measure. To fix the ideas, suppose that the hallmark to be studied is the weight of a population, and that the unity of measure is the gram. In this case, to have a homogenous population, we have to restrict it, for example, with respect to age, sex, social class and so on. 
On the basis of statistical mechanics, to construct a model able to study the evolution of some selected hallmark of the multi-agent system, the fundamental assumption is that agents are indistinguishable \cite{PT13}. This means that an agent's state at any instant of time $t\ge 0$ is completely characterized by the measure $w \ge0$ of his hallmark.
The unknown is the density (or distribution function) $f = f(w, t)$, where $w\in \R_+$ and the time $t\ge 0$. Its time evolution is described, as shown later on, by a kinetic equation
of Boltzmann type.
The precise meaning of the density $f$ is the following. Given the system of agents to study, and given an interval or a more complex sub-domain $D  \subseteq \R_+$, the integral
\[
\int_D f(w, t)\, dw
\]
represents the number of individuals with  are characterized by a measure $w \in D$ of the hallmark at time $t > 0$. It is assumed that the density function is normalized to one, that is
\[
\int_{\R_+} f(w, t)\, dw = 1.
\]
The change in time of the density is due to the fact that agents continuously upgrade the measure $w$ of their hallmark in time by some action.  To maintain the connection with classical kinetic theory of rarefied gases, we will always refer to a single upgrade of the measure as an \emph{interaction}. 
 
In the problem we want to study, the result depends on some human behavior, that can be translated by saying that agents tend to increase the value $w$ by interactions.  Referring to the example of the weight, this behavior is clearly satisfied, since it is more pleasant to eat (and to gain weight) than to fast (and to loose it). 
In reason of the existence of this human tendency, and to avoid problems related to an abnormal growth, it is fixed for the agents an upper ideal measure $\bar w$ of the hallmark relative to the homogenous class, as well as a second value  $\bar w_L$, with $\bar w_L > \bar w$, threshold value that would be better not to exceed.  Consequently, the human tendency to increase the value $w$ by interactions has to be coupled with the existence of this limit value $\bar w_L$ which would be better not to overcome.  This is the classical situation excellently described by Kahneman and Twersky  in their pioneering paper  \cite{KT}, devoted to describe decision under risk. Inspired by this idea, we will describe an agent's interaction as
 \be\label{coll}
 w_* = w  - \Phi(w/\bar w_L) w + \eta w.
 \ee
 The function $\Phi$ plays the role of the \emph{value function} in the prospect theory of Kahneman and Twersky\cite{KT}. In  \cite{KT} a classical value function   is positive and concave above the reference value $1$ ($w > \bar w_L$), while negative and convex below ($w < \bar w_L$).  At difference with the choice of Kahneman and Twersky we will assume as value function  the increasing concave function
 \be\label{vf}
 \Phi(s) = \mu \frac{s^\delta -1}{s^\delta +1} , \quad  s \ge 0.
 \ee
In \fer{vf} $0<\mu < 1$ and $0< \delta < 1$ are suitable constants characterizing the agents behavior. In particular, the value $\mu$ will denote the maximal amount of variation  that agents will be able to obtain in a single interaction. Note indeed that the value function $\Phi(s)$ is such that 
 \be\label{bounds}
  |\Phi(s)| \le \mu.
 \ee
 The function in \fer{vf} maintains most of the physical properties of the value function of prospect theory required in  \cite{KT}, and is particularly well adapted to the present situation.
The presence of the minus sign in front of the value function $\Phi$ is due to the obvious fact that an agent will respect his tendency to increase the value $w$ when $w < \bar w_L$,  while it will be induced to decrease it if $w >\bar w_L$. Note moreover that the function $\Phi(s)$ is such that, given  $0 < s < 1$
 \[
 - \Phi\left(1-s \right) > \Phi\left(1+s \right).
 \]
Therefore, given  two agents starting at the same distance from the limit value $\bar w_L$ from below and above,  the agent starting below will move closer to the optimal value, than the agent starting above. 

Last, to take into account a certain amount of human unpredictability in the outcome of an interaction, it is reasonable to assume that any result can have random variations (expressed by $\eta w$ in \fer{coll}) which in the mean are negligible, and in any case are not so significant to produce a sensible variation of the value $w$.  Also, to be consistent with the necessary positivity of the value $w_*$, it is assumed that the random variable $\eta$ can take values in the interval $(-1 +\mu, +\infty)$, while $\langle \eta\rangle = 0$. Here and after, $\langle \cdot \rangle$ denotes mathematical expectation. It will further assumed that the variance $\langle \eta^2\rangle = \lambda$, where clearly $\lambda >0$.

\begin{remark} 
Clearly, the choice of the value function  \fer{vf} is only one of the possible choices. For example, to differentiate the percentage of maximal increasing of the value from the percentage of decreasing it, and to outline the difficulty to act again the tendency, one can consider the value function
 \be\label{diff}
  \Psi(s) = \mu\frac{s^\delta -1}{\nu s^\delta +1} , \quad  s \ge 0,
 \ee
where the constant $ \nu >1$. In this case, \fer{bounds} modifies to
 \be\label{bound2}
 -\gamma \le \Psi(s) \le  \frac \mu{\nu} < \mu.
  \ee
In this case, the possibility to go against the natural tendency is slowed down. As we shall see in Section \ref{quasi}, this choice will modify the parameters of the steady state distribution.
\end{remark}
Given the \emph{interaction}  \fer{coll}, the study of the
time-evolution of the distribution of the values of the hallmark under study can be obtained by
resorting to kinetic collision-like models \cite{Cer,PT13}. The variation of the  density $f(w,t)$  obeys to a linear
Boltzmann-like equation. This equation is usually and fruitfully written
in weak form. It corresponds to say that the solution $f(w,t)$
satisfies, for all smooth functions $\varphi(w)$ (the observable quantities)
 \begin{equation}
  \label{kin-w}
 \frac{d}{dt}\int_{\R_+}\varphi(w)\,f(w,t)\,dx  = \frac 1\tau
  \Big \langle \int_{\R_+} \bigl( \varphi(w_*)-\varphi(w) \bigr) f(w,t)
\,dw \Big \rangle.
 \end{equation}
 Here expectation $\langle \cdot \rangle$ takes into account the presence of the random parameter $\eta$ in \fer{coll}. The positive constant $\tau$ measures the interaction frequency.

The right-hand side of equation \fer{kin-w} represents the difference in density between agents that modify their value from $w$ to $w_* $ (loss term with negative sign) and   agents  that  change their value from  $w_*$ to  $w$  (gain term with positive sign).

In reason of the nonlinearity (in the hallmark variable $w$) of the interaction \fer{coll}, it is immediate to verify that the only conserved quantity of equation \fer{kin-w} is obtained by setting $\varphi = 1$. This conservation law implies that the solution to \fer{kin-w} remains a probability density for all subsequent times $t >0$.  The evolution of other moments is difficult to follow. As main example, let us take $\varphi(w) = w$, which allows to obtain that the evolution of the mean value
 \[
 m(t) = \int_{\R_+}w\, f(w, t)\, dw.
  \]
Since
 \[
 \langle w_* - w \rangle = \mu\frac{w^\delta -\bar w_L^\delta}{w^\delta +\bar w_L^\delta}\, w,
  \]
we obtain
 \be\label{evo-m}
 \frac{d\,\, m(t)}{dt} = \frac\mu\tau\int_{\R_+} \frac{w^\delta -\bar w_L^\delta}{w^\delta +\bar w_L^\delta}\, w\, f(w,t)\, dw.
 \ee
Note that equation \fer{evo-m} is not explicitly solvable.  However, in view of condition \fer{bounds} the mean value of the solution to equation \fer{kin-w} remains bounded at any time $t >0$, provided that it is bounded initially, with the explicit upper bound
 \[
 m(t) \le m_0\exp \left\{\frac\mu\tau \, t \right\}.
  \]
Analogous result holds for the evolution of the second moment, which corresponds to assume $\varphi(w) = w^2$. In this case, since
 \[
 \langle w_*^2 -w^2\rangle = \big[ \Phi\left( w/\bar w_L \right)^2 -2 \Phi\left( w/\bar w_L\right) + \lambda \big] w^2 \le [\mu^2 + \lambda]w^2
  \]
the boundedness of the initial second moment implies the boundedness of the second moment of the solution at any subsequent time $t>0$, with the explicit upper bound
\[
 m_2(t) \le m_{2,0}\exp \left\{\frac{\mu^2 +\lambda}\tau \, t \right\}.
  \]
\section{Quasi-invariant  limit and the Fokker-Planck equation}\label{quasi}

The linear kinetic equation \fer{kin-w} describes the evolution of the density consequent to interactions of type \fer{coll}, and it is valid for any choice of the parameters $\delta, \mu$ and $\lambda$. In real situations, however, it happens that a single interaction (any meal in the case of weight) determines only an extremely small change of the value $w$. This situation is well-known in kinetic theory of rarefied gases, where interactions of this type are called \emph{grazing collisions} \cite{PT13,Vi}.  The presence of this smallness can be easily achieved by setting in \fer{coll} for some value $\e$, with $\e \ll 1$ 
 \be\label{scal}
\delta \to \e\delta, \quad \eta \to \sqrt\e \eta.
 \ee
This scaling allows to retain the effect of all parameters in the forthcoming limit procedure. An exhaustive discussion on these scaling assumptions can be found in  \cite{FPTT}. Using \fer{scal}, since for any time $t >0$ we can write \fer{evo-m} as
 \[
\frac{d }{dt}m(t) = \e \bar w_L \, \, \frac\mu\tau\int_{\R_+} \frac 1\e\left[ \left(\frac w{\bar w_L}\right)^{\e\delta} -1\right] \, \frac w{\bar w}\, \frac 1{\left(w/\bar w_L\right)^{\e\delta} + 1}\, f(w,t)\, dw, 
 \]
 and, for $s \ge 1$, independently of the value of the small parameter $\e$
 \[
 \frac 1{\e\delta}\left[ s^{\e\delta} -1\right]  \le s,
  \]
 while for $s \le 1$
 \[
  \frac 1{\e\delta}e\left[ s^{\e\delta} -1\right]  s \ge -1,
 \] 
the scaling \fer{scal} is such that, for any given fixed time $t >0$, the consequent variation of the mean value $m(t)$ is small with $\e$ small.  In this situation it is clear that, if we want to observe an evolution of the average value independent of $\e$,  we need to resort to a scaling of the frequency $\tau$. If we set $\tau \to \e \tau $, and $f_\e(w, t) $ will denote the density corresponding to the scaled interaction and frequency, then  the evolution of the average value for $f_\e(w, t)$ satisfies
\[
\frac{d}{dt}\int_{\R_+}w \,f_\e(w,t)\,dw  =  \bar w_L\, \, \frac{\mu\delta}\tau\int_{\R_+} \frac 1{\e\delta}\left[ \left(\frac w{\bar w_L}\right)^{\e\delta} -1\right] \, \frac w{\bar w_L}\, \frac 1{\left(w/\bar w_L\right)^{\e\delta} + 1}\, f_\e(w,t)\, dw, 
 \]
namely a computable evolution law for the average value of $f$, which remains bounded even in the limit $\e \to 0$, 
since pointwise
 \be\label{AA}
 A_\e (w) = \frac 1{\e\delta}\left[ \left(\frac w{\bar w_L}\right)^{\e\delta} -1\right] \, \frac 1{\left(w/\bar w_L\right)^{\e\delta} + 1} \to \frac 12 \log \frac w{\bar w_L}.
  \ee
 The reason behind the scaling of the frequency of interactions is clear. Since the scaled interactions  produce a very small change in the value $w$, a finite observable variation of the density can be observed in reason of the fact that each agent has a huge number of interactions  in a fixed period of time. 
By using the same scaling \fer{scal} one can easily obtain the evolution equation for  the second moment of $f_\e(w,t)$, which will be well-defined also in the limit $\e \to 0$ (cf. the analysis in  \cite{FPTT}).  

The previous discussion about moments enlightens  the main motivations and the mathematical ingredients that justify the passage from the kinetic model \fer{kin-w} to its continuous counterpart given by a Fokker--Planck type equation. 
Given a smooth function $\varphi(w)$, and a collision \fer{coll} that produces a small variation of the difference $w_*-w$, we can expand in Taylor series $\varphi(w_*)$ around $\varphi(w)$. Using the scaling \fer{scal} one obtains
 \[
\langle w_* -w \rangle = - \e\delta \, \mu \,A_\e(w)\,w;  \quad  \langle (w_* -w)^2\rangle =  \left(\e^2 \, \mu^2\, \delta^2 \,A_\e^2(w) + \e \lambda\right) w^2.
 \]
Therefore, equating powers of $\e$,  we obtain the expression
 \[
\langle \varphi(w_*) -\varphi(w) \rangle =  \e \left( - \varphi'(w)\frac {\mu\delta} 2\,w \log \frac w{\bar w_L}
  + \frac \lambda 2 \, \varphi''(w)  w^2 \right) + R_\e (w),
 \]
where the remainder term $R_\e$, for a suitable $0\le \theta \le 1$, is given by 
 \begin{equations}\label{rem}
R_\e(w) = & \frac 12 (\e\mu\delta)^2 \, \varphi''(w) A_\e^2(w)\, w^2  + \e \mu\delta \left( A_\e(w) - \frac 12 \log \frac w{\bar w_L}\right)\,w +
\\  
& \frac 16  \langle \varphi'''(w+\theta(w_* -w))  (w_* -w)^3\rangle, 
 \end{equations}
 and it is such that 
  \[
  \frac 1\e \, R_\e(w) \to 0
  \]
 as  $\e \to 0$. Therefore, if  we set  $\tau \to \e \tau$,  we obtain that the evolution of the (smooth) observable quantity $\varphi(w)$ is given by
\[
\begin{aligned}
 & \frac{d}{dt}\int_{\R_+}\varphi(w) \,f_\e(w,t)\,dw  = \\
 & \int_{\R_+} \left( - \varphi'(w)\,  \frac {\mu\delta} 2 \, w\, \log \frac w{\bar w_L} + \frac \lambda 2 \varphi''(w) w^2 \right) f_\e(w,t)\, dw \ + \frac 1\e \mathcal R_\e(w,t) ,
 \end{aligned}
 \]
where 
\[
\label{rem3}
\mathcal R_\e(t) = \int_{\R_+ } R_\e(w)  f_\e(w,t)\, dw,
\]
and $R_\e$ is given by \fer{rem}. Letting $\e \to 0$,  shows that in consequence of the scaling \fer{scal} the weak form of the kinetic model \fer{kin-w} is well approximated by the weak form of a linear Fokker--Planck equation (with variable coefficients)
\begin{equations}
  \label{m-13}
 & \frac{d}{dt}\int_{\R_+}\varphi(w) \,g(w,t)\,dw  = \\
  & \int_{\R_+} \int_{\R_+} \left( - \varphi'(w) \frac \gamma 2 \, w\, \log \frac w{\bar w_L} + \frac \lambda 2 \varphi''(w) w^2 \right) g(w,t)\, dw 
   \end{equations}
 in which we defined $\gamma = \delta\mu$.   
Provided the boundary terms produced by the integration by parts vanish,  equation \fer{m-13} coincides with the weak form of the Fokker--Planck equation
 \begin{equation}\label{FP2}
 \frac{\partial g(w,t)}{\partial t} = \frac \lambda 2 \frac{\partial^2 }{\partial w^2}
 \left(w^2 g(w,t)\right )+ \frac \gamma 2
 \frac{\partial}{\partial w}\left(  w\, \log \frac w{\bar w_L} g(w,t)\right).
 \end{equation}
Equation \fer{FP2} describes the evolution of the distribution density $g(w,t)$ of the hallmark  $w \in \R_+$, in the limit of the \emph{grazing} interactions.  As often happens with Fokker-Planck type equations, the steady state density can be explicitly evaluated, and it results to be a lognormal density, with parameters linked to the details of the microscopic interaction \fer{coll}.
 
\section{The steady state is a lognormal density}

The stationary distribution of the Fokker--Planck equation \fer{FP2} is easily found by solving the differential equation 
 \be\label{sd}
 \frac \lambda 2 \frac{d }{dw}
 \left(w^2 g(w)\right )+ 
 \frac \gamma 2  w\, \log \frac w{\bar w_L}\, g(w)   =0.
 \ee
Solving \fer{sd} with respect to $h(w)= w^2 g(w)$ by separation of variables gives as unique solution to \fer{sd} of unit mass the density
 \be\label{equilibrio}
g_\infty(w) = \frac 1{\sqrt{2\pi \sigma}\, w} 
\exp\left\{ - \frac{(\log w - \kappa)^2}{2 \sigma}\right\},
 \ee 
where 
 \be\label{para}
 \sigma = \frac \lambda\gamma,  \quad \kappa = \log \bar w_L - \sigma.
 \ee
 Hence, the equilibrium distribution \fer{equilibrio} takes the form of a lognormal density with mean and variance given respectively by
 \be\label{moments}
 m(g_\infty) = \bar w_L e^{-\sigma/2}, \quad Var(g_\infty) = \bar w_L^2 \left( 1 - e^{-\sigma}\right).
 \ee
Note that the moments are expressed in terms of the parameters $\bar w_L$, $\lambda$ and $\gamma = \delta\mu$ denoting respectively the limit value $\bar w_L$, the variance $\lambda$ of the random effects and the value $\gamma= \delta\mu$ of the value function $\phi$. Note moreover that, since $\delta<1$ in \fer{coll}, the constant $\gamma <\mu$, namely less than the maximal amount of weight that can be lost in a single interaction.

In particular, both the mean value and the variance of the steady state density depend only on the ratio $\sigma = \lambda/\gamma$  between the variance $\lambda$ of the random variation of weight allowed, and the portion $\delta\mu$ of maximal percentage $\mu$ allowed of possible variation of weight in a single interaction.

If the value $\sigma$ satisfies 
 \be\label{www}
 \sigma \ge 2 \log \frac{\bar w_L}{\bar w},
 \ee 
the mean value is lower that the fixed ideal value $\bar w$, which represents a very favorable situation for the population of agents. 

\begin{remark} 
If the value function \fer{diff} is considered, then \fer{AA} will be substituted by
 \be\label{AB}
 A_\e (w) = \frac 1\e\left[ \left(\frac w{\bar w_L}\right)^\e -1\right] \, \frac 1{\nu \left(w/\bar w_L\right)^\e + 1} \to \frac 1{1+\nu}  \log \frac w{\bar w_L}.
  \ee
 where the constant $ \nu >1$. In this case, the drift term in the Fokker--Planck equation \fer{FP2} modifies to
 \be\label{drift2}
D(g)(w) =   \frac \gamma {1+\nu}
 \frac{\partial}{\partial w}\left(  w\, \log \frac w{\bar w_L} g(w,t)\right)
   \ee
In this case, by setting
 \[
 \tilde \gamma =  \gamma \, \frac 2 {1+\nu} < \gamma,
  \]
the steady state \fer{equilibrio} remains a lognormal density, with $\sigma$ substituted by $\tilde\sigma = \sigma(1+\nu)/2 > \sigma$. 
\end{remark}

\section{Examples}\label{examples}

\subsection{Body weight distribution}\label{weight}

The current emphasis on probabilistic approaches to risk assessment makes information on the complete body weight distribution, not just the average weight, important \cite{PTR}. In addition, these distributions are needed not just for the overall population, but for different groupings, including age and sex subgroups.

Various studies based on data periodically collected on the health and nutrition status of U.S. residents by NHANES, (National Health and Nutritional
Examination Survey)  show that body weights tend to follow a log-normal distribution \cite{BB,BC}.  To confirm this observation with data
from various time different data collected by NHANES, in  \cite{PTR} both graphical and formal
goodness-of-fit tests were performed within each sex for each age group. These tests indicated that the lognormal assumption provides a reasonable fit (results not shown due to the sheer volume of analysis results).
For the most part, the log-normal distribution is adequate for each age and sex category, while the fit is poorer for data sets where a wide range of adjacent age categories are combined for the assessment. This reduced fit has been explained by age-based changes in log-body-weight mean and standard deviation.

In the case of body weight, the appearance of lognormal distribution can be fully justified on the basis of the previous kinetic modeling. Let us fix a population of agents, homogeneous with respect to age, sex and other possible classes. Agents have a precise indication about the  ideal weight  $\bar w$ of their homogenous class, through media advertisements, web and others. Also, agents perfectly know that to remain below this ideal weight $\bar w$ has a lot of advantages in terms of health. However, this bound is in conflict with the pleasure of eating. At the end, it can be reasonably assumed that agents will be really afraid of having consequences about the size of their weight only when this weight overcomes a certain limit value $\bar w_L$, with $\bar w_L > \bar w$.  Consequently,  an agent with a weight $w > \bar w_L$, will in most cases assume a reduced number of calories with the scope to have a weight reduction. On then other side, in the case in which the agent is so lucky to be under weight, expressed by a value $w < \bar w_L$, he will fully enjoy his meal, which could naturally lead to a progressive increasing  value of his weight. It is clear that the two situations are completely different, since in the former an agent will be worried about his weight, while in the latter the agent will be fully relaxed.
Therefore, given  two agents starting at the same distance from the limit weight $\bar w_L$ from below and above, it is easier (and more pleasant) for the agent starting below to move closer to the optimal weight, than for the agent starting above. Indeed, the perception  an agent will have of his weight will be completely different depending of the sign of the value function. 

Last, it is clear that there is a certain amount of unpredictability in weight variations. This random variability can be easily recognized to be consequence of diseases or dysfunctions, as well as consequence of a diffuse ignorance of the caloric content or of the glycemic index of foods.   

Hence, an interaction of type \fer{coll} fully agrees with the case of body weight. Also, the grazing limit considered in Section \ref{quasi} is highly justified. Indeed, it is clear that single interactions, which are here represented generically by meals, produce only an extremely  small variation of body weight, while sensible variations of weight can be observed only after a certain period of time (weeks or more). The evident consequence of this observation is that the Fokker--Planck equation \fer{FPori}, once the relevant parameters  are properly identified, provides a good description of the phenomenon of lognormal distribution of body weight in a homogeneous population of agents.

\subsection{The age of first marriage}

It is known that in the western countries the number of women marrying for the first time tends to be distributed quite closely lognormally by age. As documented in  \cite{Pre} the lognormal fit  was fairly good for women in the United Kingdom in 1973 (as it is for many other years and western countries).  In many situations, moreover, there is concentration of marriages  around a certain age, which is commonly hypothesized to be due to social pressure. In many cases, like the data of United Kingdom  that Preston analyzed \cite{Pre}, the data variations are difficult to interpret, since women  age at marriage was not necessarily the age at first marriage. 

The mathematical modeling of Sections \ref{model} and \ref{quasi} allows to justify the lognormal variations of the age of first marriage of women in western countries. In this case, the hallmark measured by $w$ will indicate the age of the first marriage,  and the characteristic microscopic \emph{interaction} is the change in  $w$ that results from a statistical control made at regular fixed intervals of time. The starting assumption is that woman  have a social pressure to be married around a certain age $\bar w$, in any case preferably remaining below a certain age $\bar w_L$, with $\bar w_L > \bar w$. It is reasonable to assume that most women will take the value $\bar w_L$ (instead of $\bar w$) as the significant age  of marriage to respect. Indeed, one can assume that a woman will be really looking at the necessity to be married only when her age tends to move away from above from the value $\bar w_L$.  The reason relies in the existence of the natural age bound to respect,  which is related to the desire for motherhood.   Consequently,  the number of woman that are not married at the age $w > \bar w_L$, will tend to decrease. Likewise, in the case in which a woman is not married at the age $w < \bar w_L$, she will in general enjoy her freedom, and she will retain preferable to postpone the age of first marriage. Note that the relative motivations are in agreement with the choice of the value function \fer{vf}.

Also in this situation, in reason of the human nature, one is forced to introduce a certain amount of unpredictability in the variation of the age of first marriage, which can be anticipated in the case in which a woman knew a new boyfriend, or postponed when  she lost the old one.  
Consequently an interaction of type \fer{coll} is in agreement with the variation in time of the age of  first marriage. Once again, let us discuss the grazing limit assumption leading to the analysis of Section \ref{quasi}. This assumption follows by assuming that the number of women which have been married in a unit of time (for example a day) is extremely small with respect to the whole population. Therefore, deterministic variations of the age of first marriage tend to be negligible in subsequent observation, while the period of time needed to observe a sensible variation has to be very high. According to the analysis of Section \ref{weight}, the evident consequence of this behavior is that the Fokker--Planck equation \fer{FPori}, once the relevant parameters  are properly identified, provides a good description of the phenomenon of lognormal distribution of the age of first marriage. 

\subsection{Modeling drivers in traffic}

The typical driving behavior for a vehicle on a busy road often follows well-established rules. The driver of the following vehicle will repeatedly adjust the velocity to track the leading vehicle and meanwhile keep a safe distance between the two vehicles. The following drivers may brake to increase the time headway or accelerate to decrease the time headway. However, this is not as easy as free driving, since the vehicles are moving as a queue and the spacing between each other can be small. In reason of the fact that the leading vehicle's movement is often unpredictable (at least not fully predictable), the accelerating and braking actions of the driver are often overdue. In particular, the behavior of drivers tends to be different concerning acceleration and decelerations.  On average, the absolute magnitudes of actual accelerations are typically smaller than that of actual decelerations, because accelerations are constrained by the separation distance to the leading vehicle. 

The previous discussion clarify the possible reasons behind the appearance of lognormal distributions in situations of crowded traffic. One of these situations has been detailed in  \cite{JJ}, by a precise fitting of the detailed distribution of the departure headway obtained by analyzing the video traffic data collected from various road intersections in Beijing during the years $2006$ and $2007$. The data were shown to be consistent with a certain lognormal distribution (though with different mean and variance values), respectively. This suggested intuitively that such distributions should be interpreted as the outcome of the interactions between the vehicles in the discharging queue. To verify this conjecture, the authors introduced a new car-following model, designed to simulate these interactions. In this model \cite{JJ}, drivers update their position according to a two-step rule which takes into account acceleration and deceleration rates. Results showed  consistency  between the observed empirical distribution and the simulated departure headway given by the  model.

Also in this context, the appearance of lognormal distribution can be  justified on the basis of the kinetic modeling assumptions of section \ref{model}. Let us fix a population of drivers, which behave according to the normal rules of safety. Given a certain mean speed of cars on the traffic line, drivers have a precise indication about the ideal distance $\bar w$ to maintain from the vehicle in front, to avoid possible car accidents. However, this ideal bound is in conflict with the usual rush to arrive as soon as possible. At the end, it can be reasonably assumed that agents will be really afraid of having consequences only when the minimal distance from the vehicle in front reaches a certain limit value $\bar w_L$, with $\bar w_L < \bar w$.  Consequently, when a driver recognizes to be at a distance  $w < \bar w_L$, will soon decelerate with the scope to increase his distance from the vehicle in front. On then other side, in the case in which the distance from the vehicle in front has a value $w > \bar w_L$, he will increase its speed, which could naturally lead to a reach a shorter distance from the vehicle in front. It is clear that the two situations are completely different, since, as discussed in  \cite{JJ} in the former an agent will be worried about his safety and his deceleration will be more pronounced, while in the latter the agent will be relaxed and his acceleration will be less pronounced.
Therefore, given  two drivers starting at the same distance from the limit $\bar w_L$ from below and above, the perception they  will have about the safety will be completely different depending of the sign of the value function. 

In this situation, random effect are fully justified in reason of the fact that the leading vehicle's movement is not fully predictable. 
Consequently an interaction of type \fer{coll} is in agreement with the variation of the distance. 

Last, the grazing limit assumption leading to the analysis of Section \ref{quasi} follows by considering that, in a crowded lane, most of the drivers continuously update their distance from the vehicle in front.  According to the analysis of Section \ref{quasi}, the evident consequence of this behavior is that the Fokker--Planck equation \fer{FPori}, once the relevant parameters  are properly identified, provides a good description of the distance distribution.

It is important to remark that, at difference with the situations studied before, here the sign of inequalities is reversed. In the case of body weight, an agent is relaxed when his weight is below the ideal one, while in this case a driver is relaxed when his distance from the car in front is above the ideal safety distance. To maintain the same direction, we can fix $v = 1/w$ as the hallmark to be studied. Then, the analysis of Sections \ref{model} and \ref{quasi} leads to the Fokker--Planck equation \fer{FP2}, with equilibrium distribution the lognormal density \fer{equilibrio}. 

On the other hand, it is well-known that, if a random phenomenon $X$ is lognormal with parameters $\kappa$ and $\sigma$, as given in \fer{equili}, then $1/X$ is lognormal with parameters $-\kappa$ and $\sigma$. This shows that the human behavior of drivers justifies the formation of a lognormal distribution of distances among vehicles.


\subsection{Consumption is more lognormal than income}

 The classic explanation for log normality of income is Gibrat's law \cite{Gib}, which essentially models income as an accumulation of random multiplicative shocks. In reference  \cite{BBL} a detailed critical analysis of data from the income distribution in countries including the United States and the United Kingdom revealed that the shape of income is close to, but not quite, lognormal, while the distribution of consumption is much closer to lognormal than income. 
The findings have been questioned in  \cite{BBL}, by means of an economic explanation of the reason why lognormal distribution is more adapted to consumption. The effective distribution of consumption is in any case very difficult to fit. Recent attempts claim that, while distribution of consumption is commonly approximated by the lognormal distribution,  consumption is better described by a double Pareto-lognormal distribution, a distribution which has a lognormal body with two Pareto tails and arises as the stationary distribution in dynamic general equilibrium models \cite{Toda}. 

On the basis of the analysis of the present paper, it can be easily argued that, together with economic explanations, the formation of a lognormal distribution in consumption could reasonably be a consequence of the  human tendency to prefer to spend than to earn by work. 

Let us consider a multi-agent system of consumers, which belong to a homogeneous set, represented by a fixed value of incomes, denoted by $w_0$.  In this case, the characteristic microscopic \emph{interaction} consists in the variation of the allowed consumption expenditures.  The basic assumption is that consumers  have a precise idea about the percentage of income to be devoted to expenditures, denoted by $\bar w$. However, since in general to spend money gives a great satisfaction, the barrier $\bar w$ is often exceeded, and the worry about possible consequences  for excessive consumption will begin only above a certain limit, denoted by $\bar w_L$, with $\bar w_L > \bar w$, but, to avoid the unpleasant possibility to have debts, $\bar w_L \le w_0$.  Consequently, if a consumer is in the situation to have spent a quantity $w > \bar w_L$, he will be careful about, to reduce its forthcoming forthcoming budget. Likewise, in the case in which the consumer realizes that he did not use the whole amount of money in expenditure, so that $w < \bar w_L$, he will leave leisurely, having the possibility to spend more money in the next occasion. 

In this situation, to take into account the possible risks linked to financial operations,  is it necessary to introduce a certain amount of unpredictability in the variation of consumption. 
Consequently an interaction of type \fer{coll} is in agreement with the variation of consumption. 

 The grazing limit assumption leading to the analysis of Section \ref{quasi} follows by considering that most of the consumption expenses have a value which is extremely small with respect to the budget at disposal of the consumer. Therefore, \emph{grazing} interaction prevail.   According to the analysis of Section \ref{weight}, the evident consequence of this behavior is that the Fokker--Planck equation \fer{FPori}, once the relevant parameters  are properly identified, provides a good description of the phenomenon of lognormal distribution of consumption. 

Even if these modeling assumptions are very elementary, and do not have a strong economic content, nevertheless they give a satisfactory answer to the lognormal fitting of real consumption data. Economic effects can obviously be taken into account, and it can be hypothesized that the variations in lognormal shape is consequent to the consideration of additional effects (cf. the discussion of Section \ref{city}). 

\subsection{The size of cities}\label{city}

The debate about city size distributions knew in recent years a renewed interest. While older studies, focussed only on large cities, argued that sizes follow a Pareto tailed distribution, or even follow exactly  the famous rank-size rule known as Zipf's law \cite{Zipf}, in the influential article \cite{Eech}, Eeckhout  showed that Pareto law does not hold when taking into account all the populated places of a country. 
This conclusion raises at least the important question to characterize at best the appropriate distribution for city sizes. In his model cities grow stochastically, and this growth process, in the pure form of Gibrat's law,  asymptotically generates a lognormal size distribution.
Eeckhout then shows that the lognormal distribution delivers a good fit to
actual city sizes in the US (cf. also  \cite{GZS,GRSC,PR,Ram}). As a matter of fact, even if the lognormal does not follow a power law in the upper tail and, hence, it is strictly speaking not compatible with Pareto and Zipf, the different distributions have similar properties in the upper tail and can become virtually indistinguishable.

As discussed in  \cite{BRS}, beside the specific intellectual curiosity to properly define the size distribution, there are theoretical reasons for investigating the matter, as competing models yield different implications. Indeed, while the seminal paper by Gabaix \cite{Ga99} predicts a Zipf's law, Eeckhout\cite{Eech} proposes an equilibrium theory to explain the lognormal distribution of cities. 

In the recent paper  \cite{GT2}, we used mathematical modeling analogous to the one presented in this paper to obtain a Fokker--Planck like equation for the size distribution of cities, by introducing interactions based on some migration rule among cities. In this picture of formation of city size, it was assumed that the rate of migration depends of the size of the city, and it is inversely proportional to the size itself \cite{GT2}. Then, the resulting steady state of the Fokker--Planck equation is close to  Pareto law.  

Among others, it seems indeed established that the main phenomenon leading to the formation of cities is  the tendency of inhabitants to migrate, tendency which relies in both cultural and socio-economic reasons, first and foremost the search for new and better living conditions.  As discussed in  \cite{MZ}, this is a relatively recent behavior. In very primitive times a small community (even a family)  was able to perform all necessary activities to survive, and  there was no need to aggregate with other families beyond the size of a tribe. This is no more true in modern times, where mutual cooperation and competition brings people to live together.  Clearly this tendency to aggregate does not work in a single direction, since while a larger part of population prefers to live in a big city, another part prefers to move towards smaller cities with the hope to reach a better quality of life.
Note that migration of population across cities can be justified on the basis of other motivations, including the possibility to adjust for resources \cite{BGS,GCCC}. In any case, as it happens in all social phenomena, there is a certain degree of randomness in the process, which takes into account the possibility that part of the variation in size of cities could be generated by unforeseeable reasons.

In  \cite{GT2} it was considered that each elementary variation  of the size $v$ of a city  is the result of three different contributes
 \be\label{cit}
 v^* = v -\Phi(v)v + I_E(v)z + \eta\, v.
 \ee  
In \fer{cit} the variable $z \in \R_+$ indicates the amount of population which can migrate towards a city from the environment. It is usual to assume that this value is sampled by a certain known distribution function, which characterizes the environment itself. 

The functions $\Phi(v)$ and $I_E(v)$ describe the rates of variation of the size $v$ consequent to internal (respectively external) mechanisms. 
Always maintaining migration as main phenomenon to justify the distribution of city size, let us consider the case in which people  have a precise idea about the  ideal size  $\bar w$ of city, in terms of quality of life, possibility of work and so on. This ideal size is today achieved in western countries by looking at the ranking of the most livable cities, ranking available every year through media advertisements, web and others. Coupling this with the fact that very big cities still remain attractive, it is reasonable to assume that citizens are preferably willing to migrate to another city when its size is below  a certain limit value $\bar w_L$, with $\bar w_L > \bar w$.  In conclusion, we can assign different values to the intention to migrate from a small city to a bigger one, rather than from a big city to a smaller one. 
Therefore, given  two citizen leaving in a city of size at the same distance from the limit size $\bar w_L$ from below and above, it is more probable for the citizen leaving in city of smaller size to move closer to the size $\bar w_L$, than for a citizen leaving in a city of bigger size. Indeed, the perception  a citizen  will have of his advantages will be completely different depending of the sign of the value function.  This justifies the choice of the \emph{internal} rate of variation $\Phi(\cdot)$  in the form
 \be\label{rata}
 \Phi(v) =  \lambda \frac{\left(v/\bar v\right)^\delta -1 }{\left(v/\bar v\right)^\delta + 1 },
 \ee
that, for small values of the parameter $\delta$ produces formula \fer{AA}.

Therefore, if we assume that the dominant effect in migration is given by a variation of type \fer{cit}, with a negligible contribution of the external immmigration term $I_E(v)$, in view of the analysis of Sections \ref{model} and \ref{quasi} we conclude that the size distribution of cities has the form of a lognormal distribution. 

On the other hand, reasons behind migration are very complex, and it is quite difficult to select one or other reason as dominant. This clearly justifies the fact that the kinetic interaction is a mixture of effects (and reasons), which give in the limit a distribution which can be closer to a Pareto or Zipf law, or to a mixteure of lognormal ones.  In any case, the kinetic modeling considered in this paper (or in  \cite{GT2}) is enough to clarify the coexistence of various distributions in terms of various different microscopic interactions.


\subsection{Service times in a call center}\label{service}
Call centers constitute an increasingly important part of business world, employing millions of
agents across the globe and serving as a primary customer-facing channel for firms in many
different industries. For this reason, many issues associated with human resources management, sales, and marketing have also become increasingly relevant to call center operations and associated academic research. Mathematical models are built up by taking into account statistics concerning system primitives, such as the number of agents working, the rate at
which calls arrive, the time required for a customer to be served, and the length of time customers are willing to wait on hold before they hang up the phone and abandon the queue. Outputs
are performance measures, such as the distribution of time that customers wait \emph{on hold} and the fraction of customers that abandon the queue before being served \cite{AAM,Brown}. 
A deep inside into service times in a call center with hundreds of operators was the object of a  statistical research  reported by Brown et al. in  \cite{Brown}. They noticed that the true distribution of service times was very close to lognormal, but
is not exactly lognormal. The analysis of the huge amount of data provided by the company, covering a one-month period, made evident the almost perfect fitting of the statistical distribution of service times to a lognormal one. Among others, the same phenomenon was noticed before \cite{Brown}, even if the conclusion there was that the true distribution is very close to lognormal, but is not exactly lognormal. After excluding short service times, the strong resemblance to a lognormal distribution was shown to hold in different months.

Lognormal shape of processing times has been occasionally recognized by researchers in telecommunications and psychology. Empirical results suggesting that the distribution of the logarithm of call duration is normal for individual telephone customers and a mixture of normals
for \emph{subscriber-line} groups was discussed in  \cite{Bol}. Also,  theoretical arguments to justify  the lognormal curve of reaction times using models from mathematical psychology were introduced in  \cite{Bre,UM}. 

Taking into account these attempts, in  \cite{GT17} we outlined the mathematical modeling of Sections \ref{model} and \ref{quasi} to justify the lognormal variations of service times in a call center. In this case,  the population of agents consists of call center employed, and the characteristic microscopic \emph{interaction} is the change in future time serving of any agent who concluded its work in a single operation in a certain time $w$. The starting assumption is that agents  have precise instructions from the service manager to conclude the service in a certain ideal time $\bar w$, in any case remaining below a certain limit time for the service, denoted by $\bar w_L$, with $\bar w_L > \bar w$. It is reasonable to assume that most agents will take the value $\bar w_L$ (instead of $\bar w$) as the significant time to respect. Indeed, agents will be really afraid of having consequences about their delays in serving only when the service time is above the value $\bar w_L$.  Consequently, if an agent concluded a service in a time $w > \bar w_L$, he will accelerate  to conclude its forthcoming service in a shorter time. Likewise, in the case in which the agent was so quick (or lucky) to conclude a service in a time $w < \bar w_L$, he will work  leisurely to conclude its forthcoming service in a better way, by using a longer time. 

Also in this situation, one needs to introduce a certain amount of unpredictability in any future realization of the service, which can be unexpectedly solved quickly in some case, or to present additional difficulties due on the non properly formulated request of the customer or on  accidental problems  to access the data relative to the request. 

Consequently an interaction of type \fer{coll} is in agreement with the agent's behavior in a call center. Last, let us discuss the grazing limit assumption leading to the analysis of Section \ref{quasi}. This assumption follows by assuming that any agent knows very well the work to be done to conclude a service. Therefore, deterministic variations of the service time relative to a well-known service tend to be negligible, while the number of services needed to have a sensible variation has to be very high. According to the analysis of Section \ref{weight}, the evident consequence of this behavior is that the Fokker--Planck equation \fer{FPori}, once the relevant parameters  are properly identified, provides a good description of the phenomenon of lognormal distribution of service times.

\section{Mathematical aspects of the Fokker--Planck equation}\label{trend}

In the physical literature, Fokker--Planck equations with logarithmic factors in diffusion and drift terms have been considered and studied before in  \cite{Lo,Pes}. However, the presence of the logarithm in the diffusion coefficient allows to find analytical closed-form solutions only in special cases. It is however remarkable that this type of equations revealed to be interesting in the field of econophysics, to modeling the exchange rate dynamics in a target zone \cite{Lo}. 

In social sciences and economics, the kinetic description of phenomena, extensively treated in the recent book      \cite{PT13}, rarely leads to the appearance of lognormal distributions. To our knowledge, lognormal densities have been found in  \cite{CPP} as self-similar solutions of a linear Fokker--Planck equation, with time-depending coefficients of diffusion and drift, describing the behavior of a financial market where a population of homogeneous agents can create their own portfolio between two investment alternatives, a stock and a bound. The model was closely related to the Levy--Levy--Solomon microscopic model in finance \cite{LLS,LLSb}.

More recently \cite{To3},  a kinetic description of the density $\Phi(v,t)$ of a multi-agent system of agents interacting through a linear interaction reminiscent of Gibrat's law \cite{Gib},  led in the grazing limit considered in Section \ref{quasi} to
a linear diffusion equation with non constant diffusion coefficient, given by
 \[
 \frac{\partial \Phi}{\partial t} =  \frac{\partial^2\left( v^2 \Phi\right)}{\partial v^2} ,
 \]
 well-known to people working in probability theory and finance, since it describes a particular version of the geometric Brownian motion \cite{Oks}. Also in this case, the lognormal density appears as self-similar solution of the diffusion equation.  It is interesting to remark that the analytical  
 derivation of self-similar solutions in  \cite{To3} (respectively in  \cite{CPP}) suggests that the right way to look at the mathematical analysis of the diffusion equation and to the Fokker--Planck equation with time depending coefficients, is to enlighten their relationships with the linear diffusion, and, respectively, with the classical Fokker--Planck equation.

This idea has been developed in  \cite{To4}. Applying this strategy, it is immediate to show that the study of the initial-boundary value problem for the Fokker--Planck equation \fer{FP2}, and the large-time behavior of its solution,  takes advantage of its strong connections with the classical one-dimensional Fokker--Planck equation for the one-dimensional density $f=f(v,t)$, with $v \in \R$ and $t \ge 0$ (cf.  \cite{Ch43})
 \be\label{FPcla}
\frac{\partial f}{\partial \t} =  \frac{\partial^2 f}{\partial v^2}  +\frac 1T\, \frac{\partial }{\partial v} \left( \, ( v - m)\,f\right),   
 \ee 
where $m$ and $T>0$ are suitable constants related to mean and variance of the stationary solution. Indeed, the unique steady solution of equation \fer{FPcla} of unit mass is is the  Gaussian density (the Maxwellian equilibrium)
 \be\label{Max-cla}
 M(v) = \frac 1{\sqrt{2\pi T}}\, \exp\left\{ - \frac{(v-m)^2}{2 T}\right\}.
 \ee
Let us briefly describe the main steps of the method developed in  \cite{To4}. Let $g_0(w)$ denote a probability density on $\R_+$. 
To avoid inessential difficulties,  let us suppose that both the initial datum and the corresponding solutions are smooth enough to justify computations. To study the initial-boundary value problem for equation \fer{FP2} one needs to specify the boundary condition on the boundary $w =0$. If mass conservation is imposed on the solution to equation \fer{FP2} (cf. the discussion in  \cite{FPTT}), the natural  boundary condition is given by the so-called \emph{no-flux} condition, expressed by
\be\label{bu}
\left. \frac{\partial }{\partial w}\left(w^2
 g(w,t) \right) + \frac\gamma\lambda\, w\,\log \frac w{\bar w_L}\, g(w,t) \right|_{w=0 } = 0, \quad t>0.
\ee
Therefore, at least formally, the Fokker--Planck equation \fer{FP2}, for a given initial density $g_0(w)$, and boundary conditions \fer{bu} has a solution $g(w,t)$, that, in consequence of mass conservation, remains a probability density in $\R_+$ for all times $t >0$. 

To start with, we will show that equation \fer{FP2} allows to obtain many equivalent formulations of \fer{FP2}, which contain the quotient $G(w,t)= g(w,t)/g_\infty(w)$, each of them useful for various purposes. Since the lognormal density $g_\infty$, stationary solution of equation \fer{FP2},  satisfies the differential equation \fer{sd}, which can be rewritten as
 \be\label{stazio}
\frac{\partial }{\partial w}\log\left(w^2
 g_\infty \right) = -\frac\gamma\lambda\, \frac 1w\,\log \frac w{\bar w_L},
 \ee 
we obtain
 \[
\lambda \frac{\partial }{\partial w}\left(w^2
 g \right) + \gamma \, w\,\log \frac w{\bar w_L}\,g = \lambda w^2\, g \left( \frac{\partial }{\partial w} \log(w^2 g) + \frac\gamma\lambda \,\frac 1w\,\log \frac w{\bar w_L}\right)=
 \]
 \[
\lambda w^2\, g   \left( \frac{\partial }{\partial w} \log(w^2 g)- \frac{\partial }{\partial w} \log(w^2 g_\infty) \right) =  \lambda w^2\, g   \frac{\partial }{\partial w} \log\frac g{g_\infty}= \lambda w^2 g_\infty  \frac{\partial }{\partial v}\frac g{g_\infty}.
 \]
Hence, we can write the Fokker--Planck equation \fer{FP2} in the equivalent form
 \be\label{FPalt}
  \frac{\partial g}{\partial t} = \lambda \frac{\partial }{\partial w}\left[ w^2 g \frac{\partial }{\partial w} \log G \right],
 \ee
which enlightens the role of the logarithm of  $G$, or in the form
 \be\label{FPal2}
  \frac{\partial g}{\partial t} = \lambda \frac{\partial }{\partial w}\left[ w^2 g_\infty \frac{\partial G}{\partial w} \right].
 \ee 
In particular, recalling \fer{stazio}, we can extract from \fer{FPal2} the evolution equation for $G(w,t)= g(w,t)/g_\infty(w)$. Indeed
 \[
  \frac{\partial g}{\partial t} = g_\infty  \frac{\partial G}{\partial t} = \frac\lambda{2} w^2 g_\infty \frac{\partial^2 G}{\partial v^2} + \frac\lambda{2} \frac{\partial }{\partial w} (w^2 g_\infty) \frac{\partial G}{\partial w}= 
 \]
 \[
\frac\lambda{2} w^2 g_\infty \frac{\partial^2 G }{\partial w^2}  -\frac\gamma{2} w\,\log \frac w{\bar w_L}\, g_\infty \frac{\partial G}{\partial w},  
 \]
which shows that $G = G(w,t)$ satisfies the equation
 \be\label{quo}
\frac{\partial G}{\partial t} = \frac\lambda{2} w^2 \frac{\partial^2 G }{\partial w^2}  - \frac\gamma{2}\, w\,\log \frac w{\bar w_L}\,  \frac{\partial G}{\partial w} .
 \ee
Also, the boundary condition \fer{bu} modifies accordingly.  For the two equivalent forms of the Fokker-Planck equation \fer{FP2}, given by \fer{FPalt} and \fer{FPal2} the boundary condition at $w=0$ takes the forms 
\be\label{BCalt} 
\left. \lambda \,w^2 g(w, t) \frac{\partial }{\partial w} \log G(w,t) \right|_{w=0} = 0, \quad t >0,
\ee 
and
\be\label{BCal2} 
\left. \lambda \,w^2 g_\infty(w)  \frac{\partial G(w,\t) }{\partial w} \right |_{w=0} =0, \quad t >0.
\ee 
Note that boundary condition \fer{BCal2} can be used for equation \fer{quo} as well.

Let us introduce the transformation of variables 
 \be\label{chiave}
  v= v(w) = \log w, \quad \tau(t) = \frac\lambda{2}\, t,
 \ee
that is well-defined and invertible for $w \ge 0$. In addition,  let us consider for $t \ge 0$ the new function $F= F(v, \tau)$, defined  by
 \be\label{inv}
  F(v, \tau) = G(w,t), 
 \ee
with $v,\tau$ defined as in \fer{chiave} 
Clearly it holds
 \be\label{newt}
 \frac{\partial G(w,t)}{\partial t} = \frac{\partial F(v,\t)}{\partial t}= \frac{\partial F(v,\t)}{\partial \t}\, \frac{d\t}{dt} = \frac\lambda{2}\,\frac{\partial F(v,\t)}{\partial \t}
 \ee
 while
 \be\label{der1}
  \frac{\partial G(w,t)}{\partial w} = \frac{\partial F(v,\t)}{\partial w} = \frac{\partial F(v,\t)}{\partial v} \, \frac{dv}{dw} = \frac 1w \frac{\partial F(v,\t)}{\partial v}, 
 \ee
and
 \be\label{der2}
  \frac{\partial^2 G(w,t)}{\partial w^2} =  \frac 1{w^2} \frac{\partial^2 F(v,\t)}{\partial v^2} - \frac 1{w^2} \frac{\partial F(v,\t)}{\partial v}. 
 \ee
Substituting into \fer{quo} we obtain that, if $G(w,t)$ satisfies equation \fer{quo}, in terms of the variables $v=v(w)$ and $\t = \tau(t)$, the function $F(v,\t)=G(w,t)$ satisfies the equation
 \be\label{new-quo}
 \frac{\partial F}{\partial \t} =  \frac{\partial^2 F}{\partial v^2}  - \frac{v - \kappa}\sigma\,  \frac{\partial F}{\partial v}, 
 \ee
where the constants $\sigma$ and $\kappa$ are defined as in \fer{para}. Moreover, the boundary condition \fer{BCal2} becomes
 \be\label{new-bu}
 \left.  \lambda\, \frac{\partial F(v,t)}{\partial v} f_\infty(v) \right|_{v= -\infty}= 0,
 \ee
 where $f_\infty(v)$ is the Gaussian function defined in \fer{Max-cla}, that is
 \be\label{Max}
 f_\infty(v) = \frac 1{\sqrt{2\pi\sigma}}\exp\left\{ - \frac{(v-\kappa)^2}{2\sigma}\right\}
 \ee
  of mean $\kappa$ and variance $\sigma$.  Now, let 
 $f_0(v)$ be a probability density in the whole space $\R$, and let $f(v,\t)$ be the unique solution to  the initial value problem for the classical one-dimensional Fokker--Planck equation 
 \be\label{FP}
\frac{\partial f}{\partial \t} =  \frac{\partial^2 f}{\partial v^2}  +\frac{\partial }{\partial v} \left(\frac{v - \kappa}\sigma \,f\right).   
 \ee
Then, by setting $F(v,\tau) = f(v,\t)/f_\infty(v)$, and repeating step by step the previous computations, we conclude that $F(v,\t)$ satisfies \fer{new-quo}. Hence, through equations \fer{quo} and \fer{new-quo}, which are obtained each other by means of the transformation \fer{chiave}, we established an easy-to-handle connection between the classical Fokker--Planck equation \fer{FP} and the Fokker--Planck equation with logarithmic drift \fer{FP2}. To appreciate the importance of this connection, given the initial value $g_0(w)$, $w \in \R_+$, of equation \fer{FP2}, let us fix as initial value for the Fokker--Planck equation \fer{FP} the function
 \be\label{init}
  f_0(v) = w g_0(w), \quad v= \log w. 
 \ee
Clearly, if $g_0(w)$ is a probability density in $\R_+$,  $f_0(v)$ is a probability density function in $\R$. Moreover, the boundary condition \fer{new-bu} reduces to require that the solution to the Fokker--Planck equation \fer{FP} satisfies a suitable decay at $v = -\infty$, condition that is shown to hold by assuming that the initial density $f_0$ has some moment bounded. Consequently, in view of the relationship \fer{inv}, any result for the Fokker--Planck equation \fer{FP} with initial density $f_0(v)$ translates into a result for the Fokker--Planck equation \fer{FP2} with initial density $g_0(w)$. 
 
The main fact is that Fokker--Planck equations with constant diffusion and linear drift have been extensively studied, and many mathematical results are available. The interested reader can refer to the seminal paper  \cite{OV} by Otto and Villani. In this paper,  the Fokker--Planck structure has been utilized to obtain various inequalities in sharp form, including inequalities of logarithmic Sobolev type, thus generalizing the approach in  \cite{To3,To99}. In particular, it has been made use of the form \fer{new-quo}. For the solution to the Fokker--Planck equation \fer{FP} it is well-known that 
the relative Shannon's entropy 
  \be\label{sha}
H(f(\t)/f_\infty) =  \int_\R f(v, \t) \log \frac{f(v,\t)}{f_\infty(v)} \, dv, 
  \ee
  converges exponentially fast in time towards zero \cite{CT14,OV,To3,To99}, provided it is initially bounded, and the initial density has finite moments up to order two. The result follows by studying the decay in time of the relative entropy, and using logarithmic Sobolev inequality to get a lower bound on the entropy production in terms of the relative entropy. At the end, for the solution to equation \fer{FP} one gets the bound \cite{OV,To3,To99}
 \be\label{dec7}
H(f(\t)/f_\infty) \le H(f_0/f_\infty)\exp\left\{ - \frac 2\sigma\, \t \right\}.
 \ee
Consider now that the relative entropy \fer{sha} can be rewritten as
 \be\label{sha1}
H(f(\t)/f_\infty) = \int_\R \left[F(v,\t)\log F(v,\t)\right] f_\infty(v)\, dv. 
 \ee
Hence, changing variable in the integral in the right-hand side of \fer{sha1} according to \fer{chiave}, and using \fer{inv}, one obtains the equality
 \be\label{equ3}
 H(g(t)/g_\infty) =  H(f(\t)/f_\infty) , \quad \t =  \frac\lambda{2} \, t,
 \ee
which implies, thanks to the time transformation \fer{chiave} 
\be\label{dec8}
H(g(t)/g_\infty) \le H(g_0/g_\infty)\exp\left\{ - \gamma \, t \right\}.
 \ee
It is important to outline that the boundedness of the second moment of the initial value of the Fokker--Planck equation \fer{FP}, required for the validity of the decay \fer{dec7}, translates, in view of \fer{init} into the condition
 \be\label{ini7}
 \int_{\R_+} |\log w^2|\, g_0(w) \, dw < +\infty.
 \ee
Even if this is not a strong condition on the initial density $g_0(w)$,  it implies at least that the initial value has to decay to zero at some rate as $w \to 0$.

 Using inequality \fer{dec8} one can easily pass to recover the time decay in some more standard distance. Indeed, Csiszar--Kullback inequality\cite{Csi,Kul} implies that
 \[
\left( \int_{\R_+} \left| g(w,t) - g_\infty(w)\right|\, dw \right)^2 \le 2 H(g(t)/g_\infty),
 \]
which implies exponential convergence in $L_1(\R_+)$-distance at the suboptimal rate $\gamma/2$.

It is interesting to remark that, at difference with the decay rate of the relative entropy of the Fokker-Planck equation \fer{FP}, the decay rate of the relative entropy of the Fokker--Planck equation \fer{FP2} does not depend of the parameter $\lambda$, which measures the intensity of the diffusion process. Going back to the physical meaning of the Fokker--Planck equation \fer{FP2}, which describes random variation of measured data in social and economic phenomena, it simply means that the lognormal diffusion is exponentially reached in time independently of the intensity of the random effects in the microscopic interaction. 

This exponential in time convergence towards the stationary solution also clarifies that the multi-agent system, even if subject to perturbation, quickly returns to equilibrium, and the data we observe fit the lognormal distribution.

\section{Lognormal distributions from real data}\label{numer}
The following  subsections show that the theoretical analysis presented in this paper, which leads to a Fokker--Planck type equation with a universal lognormal equilibrium density,   is in good agreement with real data  in various situations described in Section \ref{examples}. We present results of numerical fitting in some selected example, in which it was possible to extract almost complete data from the pertinent web sites: (i) the women age at the first marriage,  (ii) the distribution of the service time in a call center, and (iii) the distribution of city size. In all cases the findings are in good agreement with the theoretical modeling.

The data analysis is performed using the open source statistical software R.
The fitting of the lognormal distribution has been obtained resorting to the {\it fitdist} package.
This package provides a function that plots the empirical probability distribution functions, 
the empirical cumulative distribution functions, the quantile-quantile plots (Q-Q plot), 
and the probability-probability plots (P-P plot). We recall here that the Q-Q plot and the P-P plot constitute 
a perfect tool to visualize a qualitative goodness of fit of the model to the data.
The closer the fitted data are to the straight line $y=x$, the better is the quality of the fitting. 

In all cases in which there were need to fit a multi-nomial lognormal distribution, we made use of the {\tt mixtools} software 
package.\footnote{https://cran.r-project.org/web/packages/mixtools, last visited, July, 17th, 2018.} 
For full details on the {\tt mixtools} package we refer the interested reader to reference  \cite{mixtools}.

\subsection{Women' Age at first Marriage}
Our first example of lognormal distribution refers to the distribution of the age of women at their first marriage. We  reproduce,
with a different data set,  the lognormal distribution already observed by Preston in his pioneering paper \cite{Pre}. We use the open data which have been published  by the municipality of the city of Milan.\footnote{\url{http://dati.comune.milano.it/data set/ds138-popolazione-matrimoni-celebrati-2003-2015}} These data contains public informations about the $36\,081$ marriages celebrated during the period 2003--2015. In agreement with the analysis of Preston, we selected from all the marriages available in the data set  the marriages of women denoted by the Italian word  ``nubile'',  namely women 
that got married for the first time.
\begin{figure}[th!]
\centering
\includegraphics[width=\textwidth]{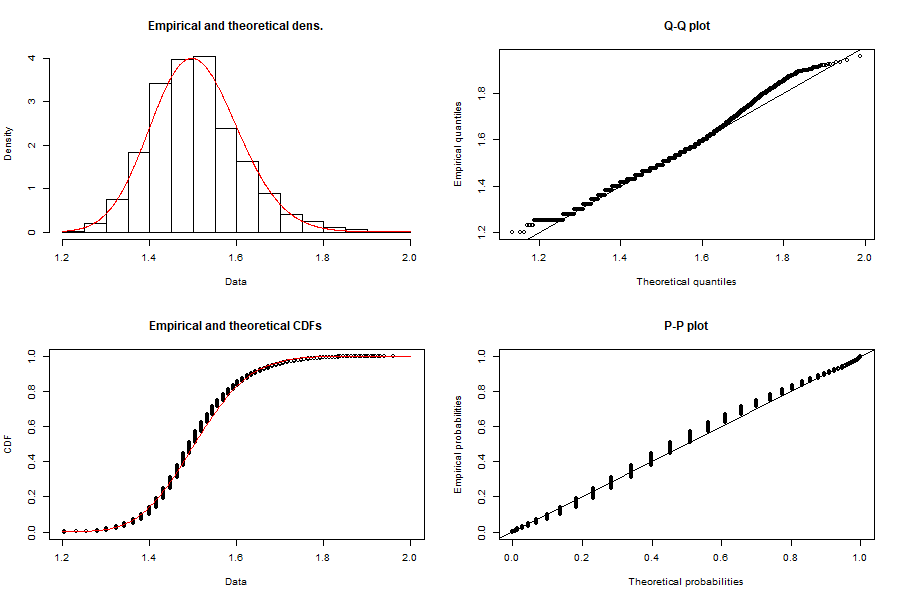}
\caption{Women' age at first marriage: Data for the city of Milan, period 2003-2016.}
\label{fig:marriage}
\end{figure}

Figure \ref{fig:marriage} shows the results obtained by fitting the data with a lognormal distribution using the {\it fitdist} package mentioned above. 
The four subplots give, in order,
 the density kernel of the fitted lognormal distribution, the empirical and theoretical cumulative distribution function,
Q-Q plot, and the P-P plot.
Note that the horizontal axis reports the exponent in log scale of the age measured in years.  
The scale is $10^x$, with $x$ ranging from 1.2 (16 years) to 1.9 (80 years).

The mean of the lognormal distribution is close to the age of 31 years, with a standard deviation of approximately 14 months.
The Q-Q plot and the P-P plot give a visual impact of the goodness of fit, which appears very good.
It is remarkable that the Q-Q plot shows a small deviation of the empirical distribution (vertical axis) from the theoretical distribution (horizontal axis), essentially due to a small number of women who get married for the first time above the age of 60 years, 
which corresponds to the value 1.8 in the plot. Overall, our results are in accordance to the results already observed in \cite{Pre} for women of United Kingdom.

\subsection{Service Times in Call Centers}
As noticed in a number of papers \cite{AAM,Brown} and recently discussed in  \cite{GT17} from the modelling point of view, lognormal distribution arises when analyzing
the distributions of the service time in a  call center. 

\begin{figure}[tph!]
\centerline{\includegraphics[width=\textwidth]{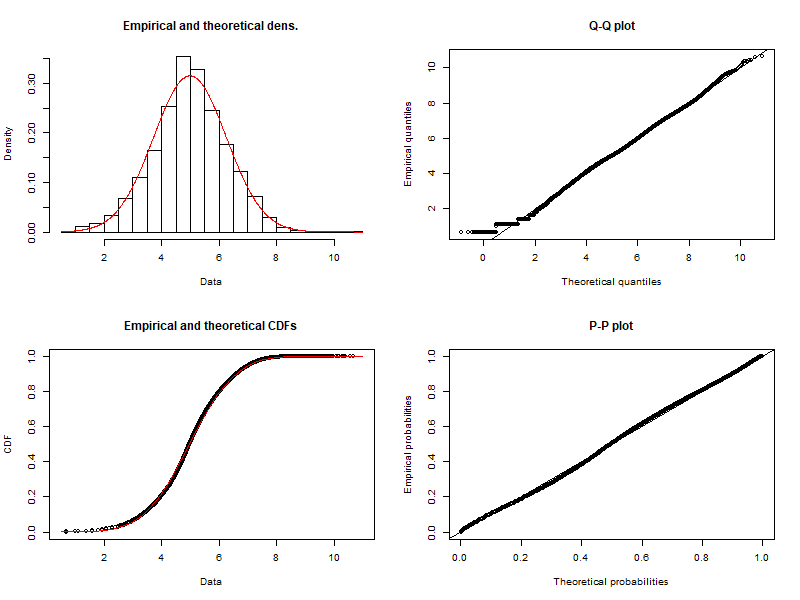}}
    \caption{Evaluation of fitting the log of the service time $log(w)$ with a Gaussian distribution.}
    \label{fig3}
\end{figure}
The analysis of reference  \cite{GT17} is relative to the call center of an Italian national communication company. 
In such a call center, every day more than 300 operators work a number of jobs
that ranges between 10\,000 and 20\,000. On the basis of the specific requests, jobs are classified into 270 different types.

Figure \ref{fig3} shows the quality of fitting with a Gaussian distribution of the logarithm of the service times, using 280'000 samples provided by the industrial partner.
As in the previous subsection, the results have been obtained using the {\it fitdist} package. The four subplots show in order:
(i) the histogram of the empirical observations along with the kernel density (red line);
(ii) the empirical and the theoretical empirical Cumulative Distribution Functions (CDF);
(iii) the Q-Q plot, and (iv) the P-P plot.
Note that in particular the Q-Q plot and the P-P plot clearly show the goodness of fitting. In both cases the data follow a straight line, with a precision even better than the one observed in Figure \ref{fig:marriage}.

Once verified that the service time distributions follows a lognormal distribution, one can perform a deeper analysis by fitting a lognormal distribution for each job type. The goal of this analysis is to observe how the distribution of the service times behaves
with respect of the ideal time $\bar{w}$, given by the service manager, and the limit $\bar{w}_L$, given by the time constraints related to the Quality of Service (QoS), which are clearly different for each job type.
As in the previous figures, the blue empirical density function refers to the real data, while the red probability density function shows the lognormal distribution with  mean $\mu$ and deviation $\sigma$ detailed in each sub-figure. 
In addition, the plots show with vertical lines the values of $\log{(\bar{w})}$ and $\log{(\bar{w}_L)}$.

For instance, the first subplot, which refers to Job Type 1, illustrates the distribution of 28\,425 service times. The blue dotted vertical line refers
to the log of the ideal time $\log{(\bar{w})=6}$ (i.e., 400 seconds), and 
the dashed green vertical line refer to the log of the time limit $\log{(\bar{w}_L)=7.3}$ (i.e., 1'500 seconds).
The fitted lognormal distribution has mean $\mu=4.9$ and variance $\sigma=1.2$.

\begin{figure}[tph!]
\centerline{\includegraphics[height=0.8\textheight]{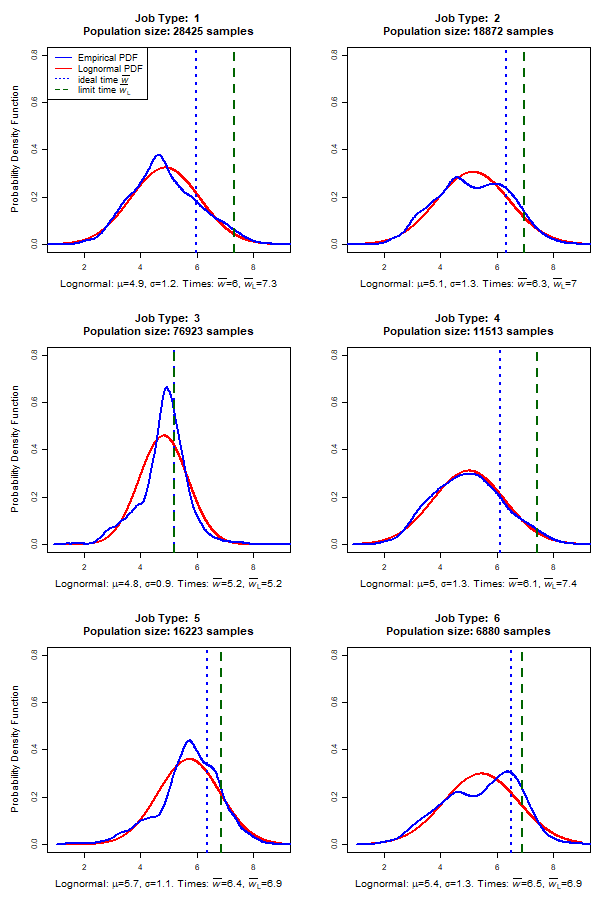}}
    \caption{Fitted lognormal distributions over the six most frequent job types. The vertical dashed lines illustrate the reference times $\bar{w}_L$ for the Quality of Service constraint.}
    \label{fig4}
\end{figure}

\begin{figure}[tph!]
\centerline{\includegraphics[width=0.7\linewidth]{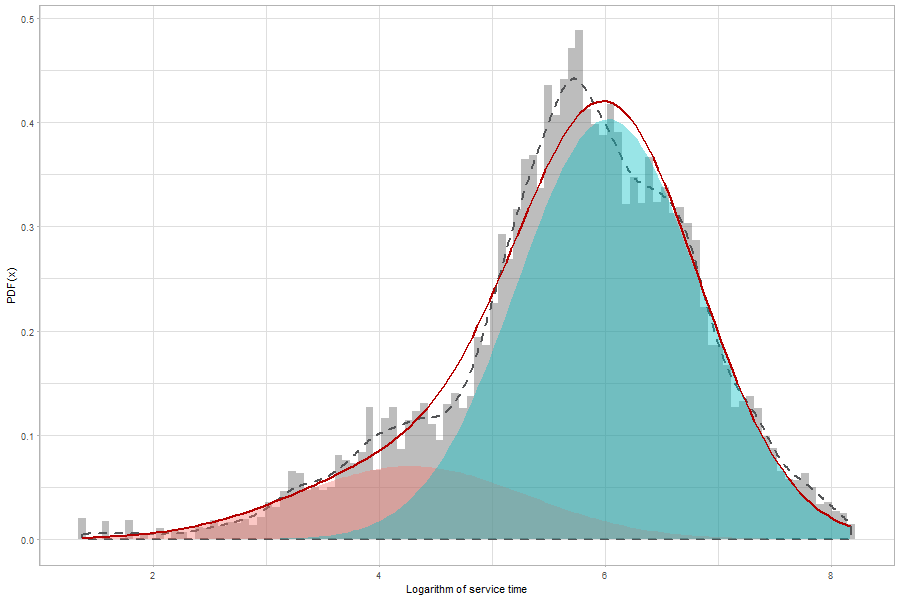}}
    \caption{Bimodal lognormal distribution: the red line is the bimodal fitted distribution, the dashed black line
    represent the kernel density, while the two red and blue areas represents the two modes of the bimodal distribution.}
    \label{fig5}
\end{figure}

Figure \ref{fig4} also shows that the service time of some job is not perfectly described by a simple lognormal distribution. For example, this can be observed for types 5 and 6. A satisfactory answer here comes from a multi model data analysis, with the objective to investigate multi modal lognormal distribution.
Figure \ref{fig5} shows the fitting of the data used for the last subplot of Figure \ref{fig4}, which corresponds to job type 6,
with a bimodal lognormal distribution. We remark that this bimodal model is able to capture a behavior of the call center operators
which tend to have two ways of working out a job: either to reject the job in a short time (represented by the first mode with
average of 30 seconds,  given by the read area in Figure \ref{fig5}), or to accept to work ``hard'' on the job and, in consequence of this decision, to conclude the service  in a longer time, with an average of more than 10 minutes.

We highlight once more the good fit of the lognormal distribution also when it is used as basic kernel in multi modal data fitting.
By using only a mixture of two lognormal distribution we perfectly catch the human behavior also in this rather complex situation. We remark that we have
analyzed numerous tests by using different models and different kernels, or employing a large number of modes. In all cases, 
by using ``only'' two lognormal distributions we got the simplest and more robust fit.

\subsection{Size Distribution of Cities}
Our last example is concerned with the problem of understanding at best the size distribution of cities \cite{GT2}.
The results that follow are extracted from  the open data published by the Italian National Institute of Statistics\footnote{http://www.istat.it, last visited June, 20th, 2018.}
and by the Swiss Federal Statistical Office.\footnote{http://www.bfs.admin.ch, last visited June, 20th, 2018.}
The first data set contains the size distribution of $8\,006$ Italian cities, ranging from the smallest village to the largest city, 
that is Rome with $2\,873\,494$ citizens. These data refer to the last official Italian census of 2016.
The second data set enumerates the size distribution of $2\,289$ Swiss cities, from the smallest one to the largest city 
that is Zurich with $396\,955$ citizens. This second data refers to the last official Swiss census of 2014.
Table \ref{tab:summary} reports the basic statistics of the two data sets, giving in order the minimum, the first quartile,
the median, the mean, the third quartile and the maximum values of city size.
Clearly, the basic statistics are clueless about the real distribution of cities size.

\begin{table}[th!]
\centering
\caption{Basic statistics of Italian and Swiss distribution of cities size.\label{tab:summary}}
\begin{tabular}{lrrrrrr}
            & Min & 1st Quart. & Median & Mean & 3rd Quart. & Max \\
\hline
Italy       & 30  & $1\,019$     & $2\,452$ & $7\,571$ & $6\,218$ & $2\,873\,494$ \\
Switzerland & 13  & 642        & $1\,425$  & $3\,638$ & $3\,513$       & $396\,955$ \\
\end{tabular}
\end{table}

\begin{figure}[th!]
\centering
\includegraphics[width=\textwidth]{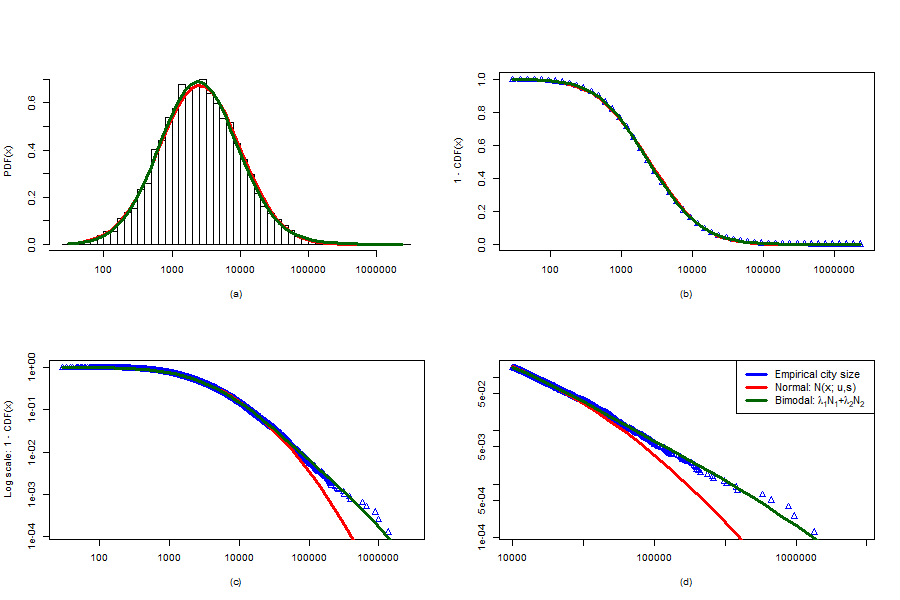}
\caption{Probability distribution function and inverse cumulative functions of Italian cities.}
\label{fig:1}
\end{figure}

\begin{figure}[th!]
\centerline{\includegraphics[width=\textwidth]{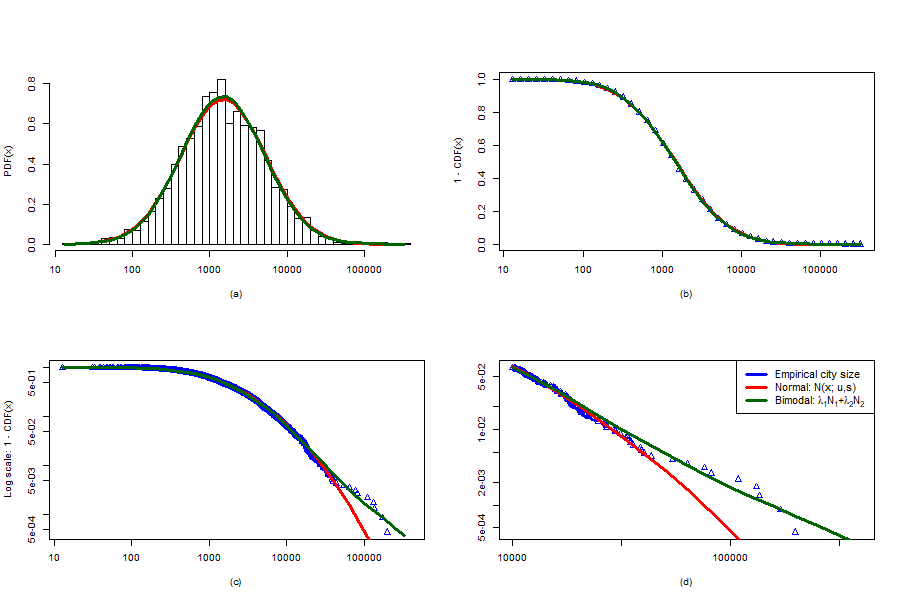}}
    \caption{Probability distribution function and inverse cumulative functions of Swiss cities.}
    \label{fig:2}
\end{figure}

In the literature, data set on distribution of the size of cities are usually studied and fitted using a Zipf's law \cite{Ga99,GCCC}
However, if we just take the logarithm of every city size and we plot the resulting distribution,
we get what looks like a classical Gaussian distribution. This can be verified through the examination of Figures \ref{fig:1}(a) and \ref{fig:2}(a), which refer respectively to the Italian and Swiss data sets. In addition, and surprisingly, it is almost impossible to distinguish between the shape of the two distributions.
Even if we looks to the inverse cumulative functions, plotted in Figures \ref{fig:2}(b) and \ref{fig:2}(b), it is pretty hard
to distinguish the resulting function from a Gaussian cumulative function.
However, if we analyze the inverse cumulative functions with  bi-logarithm plots, it is possible to notice
that a single Gaussian does not capture the trend of the tails of the distribution. This appears evident by looking
at the red lines in Figures \ref{fig:1}(c), \ref{fig:1}(d), \ref{fig:2}(c), and \ref{fig:2}(d).
On the contrary, a single Gaussian is able to perfectly fit the lower tails, which are never captured by the celebrated Zipf's law.

\begin{figure}[th!]
\centering
\includegraphics[width=0.6\textwidth]{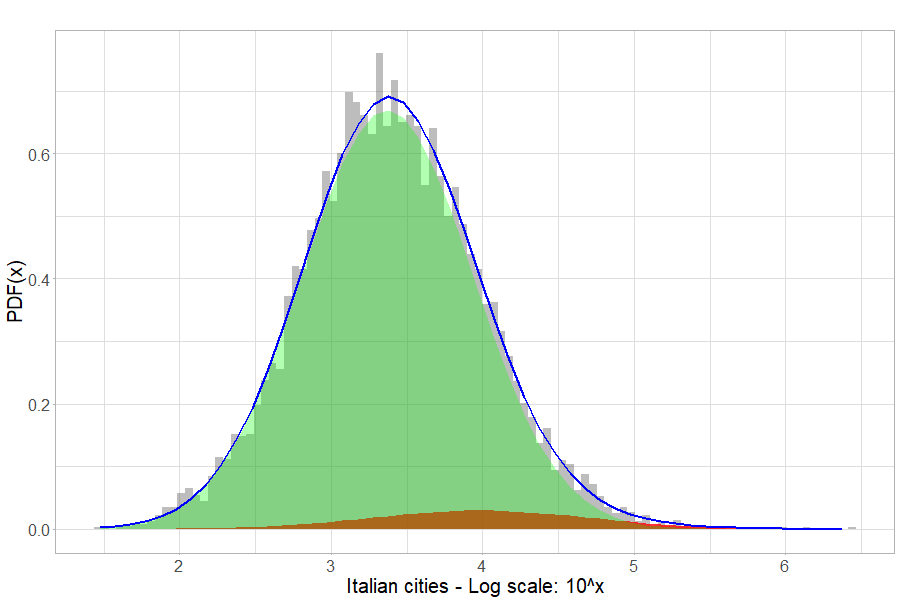}
\caption{Log Size Distribution of 8006 Italian Cities, Census 2016.} \label{fig:3}
\end{figure}

\begin{figure}[th!]
\centerline{\includegraphics[width=0.6\textwidth]{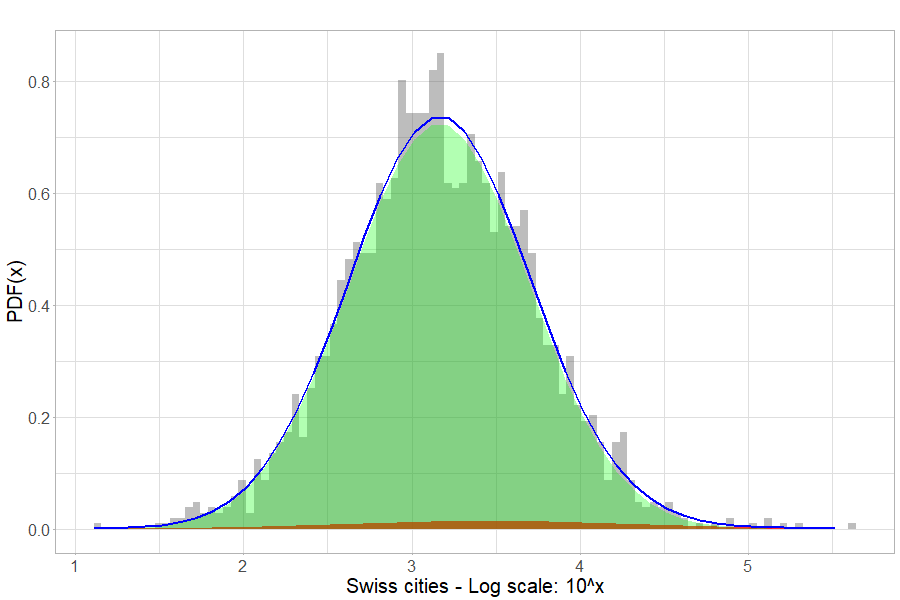}}
    \caption{Log Size Distribution of 2289 Swiss Cities, Census 2014.}
    \label{fig:4}
\end{figure}

In order to improve the fitting of the distributions also on the higher tails, it is enough to fit the
distributions of cities sizes using a multi-modal Gaussian model, by resorting to the {\tt mixtools} software package,\footnote{https://cran.r-project.org/web/packages/mixtools, last visited, June, 20th, 2018.} 
available in the R statistical programming language. For full details on the {\tt mixtools} package we refer the interested reader to [Z].
Basically, using {\tt mixtools} one is able to fit the distribution of city size with a mixture of only two Gaussians
\begin{equation}\label{eq:bimodal}
g(x) = \lambda_1 N(x; \mu_1, \sigma_1) +  \lambda_2 N(x; \mu_2, \sigma_2)
\end{equation}
Table \ref{tab:bimodal} reports the parameters fitted by {\tt mixtools} for both data sets and Figures \ref{fig:3} and \ref{fig:4} show
the respective probability density functions. It is evident that for both data sets there is a \emph{dominating} Gaussian,
since $\lambda_1 = 0.945$ for Italy and $\lambda_1 = 0.967$ for Switzerland.
In addition, there are the two tiny Gaussians (characterized by the small values of $\lambda_2$) that capture the behavior of the higher tails,
and which have both larger means and larger deviations. We remark that the blue solid line on top of the histograms
represents the corresponding bimodal distribution.
Finally, by looking at the green lines in Figures \ref{fig:1}(c), \ref{fig:1}(d), \ref{fig:2}(c), and \ref{fig:2}(d),
the goodness of fitting cities size distributions with a mixture of two Gaussian is striking evident.

\begin{table}[th!]
\centering
\caption{Mixture of two Gaussians: Model Parameters.\label{tab:bimodal}}
\begin{tabular}{lrrrrrr}
            & $\lambda_1$ & $\mu_1$ & $\sigma_1$ & $\lambda_2$ & $\mu_2$ & $\sigma_2$ \\
\hline
Italy       & 0.945 & 3.371 & 0.563 & 0.054 & 3.993 & 0.731 \\
Switzerland & 0.967 & 3.162 & 0.533 & 0.032 & 3.483 & 0.896 \\
\end{tabular}
\end{table}

\section{Conclusions}

We introduced and discussed in the present paper a number of social and economic phenomena which are characterized by data which present a good fitting with the lognormal distribution. We tried to explain this common behavior in terms of human behavior. In all cases, agents want to achieve a well-defined goal, characterized by a certain fixed value, while the rate of change in their approach is different, and  depends on the side from which an agent is looking at the fixed value.  The kinetic modeling is based on microscopic variations of the quantity under consideration which are obtained by resorting to a strong analogy with the arguments of prospect theory \cite{KT,KT1} suitable to model an agent's decision under risk. Well-known arguments of kinetic theory then allow to model these phenomena by means of a Fokker--Planck equation with variable coefficients of diffusion and drift, which exhibits a lognormal density at equilibrium. Interestingly enough, this Fokker--Planck equation can be exhaustively studied from the mathematical point of view, since it results to be linked to the classical Fokker--Planck equation, for which a number of results are available in the pertinent literature.  

It is clear that the examples considered in this paper cover only partially the huge number of phenomena in which human activity is mainly subject to a skewed perspective. Also, the numerical evidence of the appearing of the lognormal distribution in these phenomena is not restricted to the few cases treated here. In any case, we retain that our analysis can constitute a reasonable and easily generalizable modeling approach to the lognormal word of the human behavior.

\vskip 0,5cm
\section*{Acknowledgement} This work has been written within the
activities of GNFM group  of INdAM (National Institute of
High Mathematics), and partially supported by  MIUR project ``Optimal mass
transportation, geometrical and functional inequalities with applications''.

This research was partially supported by the Italian Ministry of Education, University and Research (MIUR): 
Dipartimenti di Eccellenza Program (2018--2022) - Dept. of Mathematics ``F. Casorati'', University of Pavia. 

\vskip 1,5cm


\begin{thebibliography}{10}

\bibitem{Ahr}
L.H. Ahrens, The log-normal distribution of the elements (A fundamental
law of geochemistry and its subsidiary), \emph{Geochimica et Cosmochimica
Acta} \textbf{5}  (1954) 49--73.

\bibitem{Aic}
J. Aitchison and J.A.C. Brown, \emph{The Log-normal Distribution},
Cambridge University Press, Cambridge, UK 1957. 

\bibitem{AAM}
Z. Aksin, M. Armony and V. Mehrotra, The modern call center: a multi-
disciplinary perspective on operations management research, \emph{Production And Operations Management POMS} \textbf{16} (6) (2007)  665--688.



\bibitem{BBL}
E. Battistin, R. Blundell and A. Lewbel,
Why is consumption more log normal than income? Gibrat's law revisited,
\emph{Journal of Political Economy} \textbf{117} (6) (2009) 1140--1154.

\bibitem{BGS}
P. Beaudry, D.A. Green and B.M. Sand, Spatial equilibrium with unemployment and wage bargaining: Theory and estimation, \emph{J. Urban Econ.} \textbf{79} (2014) 2--19.

\bibitem{BRS}
M. Bee, M. Riccaboni and S. Schiavo, The size distribution of US cities: not Pareto, even in the tail, \emph{Economics Letters}
\textbf{120} (2013) 232--237.

\bibitem{BHT}
N. Bellomo, M. A. Herrero and A. Tosin, On the dynamics of social conflicts looking
for the Black Swan, \emph{Kinet. Relat. Models} (6)(2013) 459-479.

\bibitem{BKS}
N.~Bellomo, D.~Knopoff and J.~Soler,
\newblock On the difficult interplay between life, complexity, and
  mathematical sciences,
\newblock {\em Math. Models Methods Appl. Sci.}  {\bf 23} (2013) 1861--1913.

\bibitem{BCKS}
N. Bellomo, F. Colasuonno, D. Knopoff and J. Soler, From a systems theory of
sociology to modeling the onset and evolution of criminality, \emph{Netw. Heterog. Media}
\textbf{10} (2015) 421--441.

\bibitem{mixtools}
T.~Benaglia, D.~Chauveau, D.~Hunter and D.~Young,  Mixtools: An R package for analyzing finite mixture models,
  \emph{Journal of Statistical Software} \textbf{32} (6) (2009) 1--29.

\bibitem{BN2}
 E. Ben-Naim, P.L. Krapivski and S. Redner, Bifurcations and patterns in compromise processes, \emph{Physica D} \textbf{183}  (2003) 190--204.

\bibitem{BN3}
E. Ben-Naim, P.L. Krapivski, R. Vazquez and S. Redner, Unity and discord in opinion dynamics, \emph{Physica A} \textbf{330}
 (2003) 99--106.

\bibitem{BN1}
E. Ben-Naim, Opinion dynamics: rise and fall of political parties, \emph{Europhys. Lett.} \textbf{69} (2005) 671--677.

\bibitem{BeDe}
M.L.  Bertotti and  M. Delitala,  On a discrete generalized kinetic approach for mo\-del\-ling persuader's influence in opinion formation processes,  \emph{Math. Comp. Model.} \textbf{48} 
(2008) 1107--1121.

\bibitem{Bol}
V. Bolotin,  Telephone circuit holding time distributions,  In
\emph{14th International Tele-Traffic Conference (ITC-14)}, 125--134, Elsevier, Amsterdam,
 (1994).

\bibitem{Bou}
L. Boudin and F. Salvarani,  The quasi-invariant limit for a kinetic model of sociological collective behavior, \emph{Kinetic Rel. Mod.} \textbf{2} (2009) 433--449.

\bibitem{Bou1}
L. Boudin and F. Salvarani, A kinetic approach to the study of opinion formation, \emph{ESAIM: Math. Mod. Num. Anal.} \textbf{43} 
(2009)  507--522.

\bibitem{Bou2}
L. Boudin, A. Mercier and  F. Salvarani, Conciliatory and contradictory dynamics in opinion formation, \emph{Physica A} \textbf{391} (2012) 5672--5684.

\bibitem{BB}
J. Brainard and D.E. Burmaster,  Bivariate distributions for
height and weight of men and women in the United States,
\emph{Risk Analysis} \textbf{12} (1992) 267--275.

\bibitem{Bre}
G. Breukelen, Theoretical note: parallel information processing models
compatible with lognormally distributed response times,  \emph{Journal of
Mathematical Psychology} \textbf{39} (1995) 396--399.

\bibitem{Brown}
L. Brown, N. Gans, A. Mandelbaum, A. Sakov,
H. Shen, S. Zeltyn and L. Zhao,
Statistical analysis of a telephone call center:
a queueing-science perspective, \emph{Journal of the American Statistical Association}
\textbf{100}, No. 469, Applications and Case Studies, March 2005.

\bibitem{BC}
D.E. Burmaster and E.A. Crouch,
Lognormal distributions for body weight as a function of age for males and females in the United States, 1976--1980,
\emph{Risk  Anal.} \textbf{17} (4) (1997) 499--505.

\bibitem{CT14}
J.A. Carrillo and G. Toscani, Renyi entropy and improved equilibration rates to self-similarity for nonlinear diffusion equations, \emph{Nonlinearity} \textbf{27} (2014) 3159--3177.



 \bibitem{CFL}
C. Castellano, S. Fortunato and V. Loreto, Statistical physics of social dynamics, \emph{Rev. Mod. Phys.} \textbf{81}
(2009) 591--646. 

\bibitem{Cer}
C. Cercignani,
\emph{The Boltzmann equation and its applications},
\newblock  Springer Series in Applied Mathematical Sciences,
  Vol.\textbf{67} Springer--Verlag, New York 1988.
  
\bibitem{ChaCha00} 
A.~Chakraborti and B.K.~Chakrabarti, Statistical
  Mechanics of Money: Effects of Saving Propensity, {\em Eur. Phys. J. B}
  \textbf{17} (2000)   167--170.  

\bibitem{Ch43}
S. Chandrasekhar, Stochastic problems in physics and astronomy, \emph{Rev. Modern Phys.} \textbf{15} (1943) 1--110.  

\bibitem{CCM} 
A. Chatterjee, B.K. Chakrabarti and S.S. Manna,
   Pareto law in a kinetic model of market with random saving propensity,
  \emph{Physica A\/} {\bf 335} (2004) 155--163.

\bibitem{ChChSt05} 
A.~Chatterjee, B.K.~Chakrabarti and
  R.B.~Stinchcombe, Master equation for a kinetic model of trading
  market and its analytic solution, {\em Phys. Rev. E} \textbf{72}   (2005) 026126.


  \bibitem{CDT}
V. Comincioli, L. Della Croce and G. Toscani,  A Boltzmann-like equation for choice formation, \emph{Kinetic Rel. Mod.} \textbf{2} (2009)  135--149.

\bibitem{CPP}
S. Cordier, L. Pareschi and C. Piatecki, Mesoscopic modelling of financial markets, \emph{J. Stat. Phys.} \textbf{134} (1) (2009)  161--184.

\bibitem{CoPaTo05} 
S.~Cordier, L.~Pareschi and G.~Toscani, On a kinetic
  model for a simple market economy, {\em J. Stat. Phys.} \textbf{120} (2005)  253--277.
    

\bibitem{Cro}
E.L. Crow and K. Shimizu eds., \emph{Log-normal distributions: theory and application}.
Marcel Dekker, New York NY 1988.

\bibitem{Csi}
I. Csiszar, Information-type measures of difference of probability distributions
and indirect observations, \emph{Stud. Sci. Math. Hung.} \textbf{2} (1967) 299-318.

\bibitem{DY00}
A.~Dr\v{a}gulescu and V.M.~Yakovenko, {Statistical mechanics of money}, {\em Eur. Phys. Jour. B} \textbf{17} (2000)  723--729.

\bibitem{DMPW}
B.~D{\"u}ring, P.A. Markowich, J-F. Pietschmann and M-T. Wolfram,
\newblock Boltzmann and {F}okker-{P}lanck equations modelling opinion formation
  in the presence of strong leaders,
\newblock {\em Proc. R. Soc. Lond. Ser. A Math. Phys. Eng. Sci.}
  {\bf 465} (2009)  3687--3708.

\bibitem{DMTb} B.\ D\"uring, D.\ Matthes and G.\ Toscani, Kinetic equations modelling wealth redistribution:
a comparison of approaches,  \emph{Phys.\ Rev.\ E} \textbf{78}   (2008) 056103.

\bibitem{Eech}
J. Eeckhout,  Gibrat's law for (all) cities, \emph{American Economic Review} \textbf{94} (2004) 1429--1451.


\bibitem{FPTT}
G. Furioli, A. Pulvirenti, E. Terraneo and G. Toscani, Fokker--Planck equations in the modelling of socio-economic phenomena, \emph{Math. Mod. Meth. Appl. Scie.} \textbf{27} (1) (2017) 115--158.

\bibitem{Ga99}
X. Gabaix, Zipf's law for cities: an explanation, \emph{Quart. J. Econom.} \textbf{114} (1999) 739--767.

\bibitem{GGS}
{S. Galam, Y. Gefen and Y. Shapir}, Sociophysics: A new approach of sociological collective behavior. I. Mean-behaviour description of a strike, {\em J. Math. Sociology} \textbf{9} (1982) 1--13.

\bibitem{GM}
S. Galam and S. Moscovici, Towards a theory of collective phenomena: consensus and attitude changes in groups, \emph{Euro. J. Social Psychology} \textbf{21} (1991) 49--74.

\bibitem{Gal}
S. Galam, Rational group decision making: A random field Ising model at $T= 0$. \emph{Physica A} \textbf{238}  (1997) 66--80.

\bibitem{GZ}
S. Galam and J.D.  Zucker, From individual choice to group decision-making. \emph{Physica A}  \textbf{287} (2000) 644--659.

\bibitem{GSV} U. Garibaldi, E. Scalas and P. Viarengo, Statistical equilibrium
in simple exchange games II. The redistribution game, \emph{Eur. Phys. Jour. B}  \textbf{60}(2) (2007)   241--246.

\bibitem{GCCC}
A. Ghosh, A. Chatterjee, A.S. Chakrabarti and B.K. Chakrabarti,
Zipf's law in city size from a resource utilization model,
\emph{Phys. Rev. E} \textbf{90} (2014) 042815.

\bibitem{Gib}
R. Gibrat,  \emph{Les inegalites economiques},  Librairie du Recueil Sirey, Paris 1931.

\bibitem{GZS}
K. Giesen, A. Zimmermann and J. Suedekum,
The size distribution across all cities-Double Pareto lognormal strikes, \emph{Journal of Urban Economics} \textbf{68} (2010) 129-137.

\bibitem{GRSC}
R. Gonz\'alez--Val, A. Ramos,  F. Sanz--Gracia and   M. Vera--Cabello,
Size distributions for all cities: Which one is best?, \emph{Papers in Regional Sciences} 
\textbf{94}  (2015) 177--196.

\bibitem{GT-ec}
S. Gualandi and G. Toscani, Pareto tails in socio-economic phenomena: a kinetic description. \emph{Economics: The Open-Access, Open-Assessment E-Journal}, 12 (2018-31): 1--17. 
 
 \bibitem{GT17}
S. Gualandi and G. Toscani, Call center service times are lognormal. A Fokker--Planck description. \emph{Math. Mod. Meth. Appl. Scie.}  \textbf{28} (8) (2018) 1513--1527.
\bibitem{GT2} 
S. Gualandi, and G. Toscani, The size distribution of cities: a kinetic explanation. Available at http://arxiv.org/abs/1807.00496 (2018).

\bibitem{Hir}
S.S. Hirano, E.V. Nordheim, D.C. Arny and C.D. Upper, Log-normal distribution
of epiphytic bacterial populations on leaf surfaces, \emph{Applied and Environmental
Microbiology} \textbf{44} (1982) 695--700. 

\bibitem{JJ}
X. Jin, Y. Zhang, F. Wang, L. Li, D. Yao, Y. Su and Z. Wei, Departure headways at signalized intersections: A log-normal distribution model approach, \emph{Transportation Research Part C} \textbf{17} (2009) 318--327.

\bibitem{Kac59}
M. Kac, \emph{Probability and related topics in physical sciences},
With special lectures by G. E. Uhlenbeck, A. R. Hibbs, and B. van der Pol.
Lectures in Applied Mathematics. Proceedings of the Summer Seminar, Boulder,
Colorado. Interscience Publishers, London-New York, 1959.



\bibitem{KT}
D. Kahneman and A. Tversky, Prospect theory: an analysis of decision under risk,
\emph{Econometrica} \textbf{47} (2) (1979) 263--292.


\bibitem{KT1}
D. Kahneman and A. Tversky, \emph{Choices, values, and frames}, Cambridge University Press, Cambridge, UK 2000.



\bibitem{Kon}
K. Kondo, The log-normal distribution of the incubation time of exogenous
diseases, \emph{Japanese Journal of Human Genetics} \textbf{21} (1977) 217--237.

\bibitem{Kul} 
S. Kullback, A lower bound for discrimination information in terms of variation,
\emph{IEEE Trans. Inf. The.} \textbf{4} (1967) 126--127.

\bibitem{LLS}
M. Levy, H. Levy and S. Solomon, A microscopic model of the stock market:
Cycles, booms and crashes, \emph{Econ. Lett.}  \textbf{45} (1994) 103--111.

\bibitem{LLSb}
M. Levy, H. Levy and S. Solomon, {\it Microscopic simulation of financial markets: from
investor behaviour to market phenomena}, Academic Press,  San Diego, CA 2000.


\bibitem{Lim}
E. Limpert, W.A. Stahel and M. Abbt,
Log-normal distributions across the sciences: keys and clues, \emph{BioScience} \textbf{51} (5) (2001) 341--352.

\bibitem{Lo}
C.F. Lo, Dynamics of Fokker--Planck equation with logarithmic coefficients and its
application in econophysics, \emph{Chin. Phys. Lett.} \textbf{27} (8) (2010) 080503.

\bibitem{Lop}
J.E. Loper, T.V. Suslow and M.N. Schroth, Log-normal distribution of bacterial
populations in the rhizosphere, \emph{Phytopathology} \textbf{74} (1984) 1454--1460.

\bibitem{LMa}
T. Lux and M. Marchesi, Volatility clustering in financial markets: a
microscopic simulation of interacting agents, \emph{International Journal
of Theoretical and Applied Finance} \textbf{3} (2000) 675--702.

\bibitem{LMb}
T. Lux and M. Marchesi, Scaling and criticality in a stocastich multi-agent
model of a financial market, \emph{Nature} \textbf{397} (11) (1999) 498--500.

\bibitem{Mal}
A. Malanca, L. Gaidolfi, V. Pessina and G. Dallara, Distribution of 226-Ra,
232-Th, and 40-K in soils of Rio Grande do Norte (Brazil), \emph{Journal of Environmental
Radioactivity} \textbf{30} (1996) 55--67.

\bibitem{MD}
D. Maldarella and L. Pareschi, Kinetic models for socio--economic dynamics
of speculative markets, \emph{Physica A} \textbf{391} (2012) 715--730.

\bibitem{MZ}
M. Marsili and Yi-Cheng Zhang, Interacting individuals leading to Zipf's Law,
\emph{Phys. Rev. Lett.} \textbf{80} (1988) 2741--2744.

\bibitem{NPT}
G.Naldi, L.Pareschi and G.Toscani eds., \emph{Mathematical modeling of
collective behavior in socio-economic and life sciences},  Birkhauser,
Boston 2010.

\bibitem{Oks}
B. Oksendal, \emph{Stochastic differential equations. an introduction with applications}, Springer-Verlag, Heidelberg 2013.

\bibitem{OV}
F. Otto and C. Villani, Generalization of an inequality by Talagrand and links with the logarithmic Sobolev inequality, \emph{J. Funct. Anal.} \textbf{173} (2000) 361--400.
  
\bibitem{PT13}
L.\ Pareschi and G.\ Toscani, \emph{Interacting multiagent systems: kinetic equations and Monte Carlo methods}, Oxford University Press, Oxford 2014.  


\bibitem{Pes}
K. Pesz, 
A class of Fokker-Planck equations with logarithmic factors in diffusion and drift terms,
\emph{Journal of Physics A} \textbf{35} (8) (2002) 1827--1832.


\bibitem{PTR}
K. Portier, J.K. Tolson and S.M. Roberts, Body weight distributions for risk assessment,
\emph{Risk Analysis} \textbf{27} (1) (2007) 11--26.

\bibitem{Pre}
 F.W. Preston,
Pseudo-lognormal distributions,
\emph{Ecology} \textbf{62} (2) (1981) 355--364.

\bibitem{PR}
M. Puente-Ajovín and A. Ramos,
On the parametric description of the French, German, Italian and Spanish city size distributions.
\emph{Ann. Reg. Sci.}  \textbf{54} (2015) 489--509.

\bibitem{Ram}
A. Ramos, 
Are the log-growth rates of city sizes distributed normally? Empirical evidence for the USA.
\emph{Empir. Econ.} \textbf{53}  (2017) 1109--1123.

\bibitem{Raz}
N.K. Razumovsky, Distribution of metal values in ore deposits, \emph{Comptes
Rendus (Doklady) de l'Acad\'emie des Sciences de l'URSS} \textbf{9} (1940) 814--816.

\bibitem{Sar1}
P.E. Sartwell, The distribution of incubation periods of infectious disease,
\emph{American Journal of Hygiene} \textbf{51} (1950) 310--318.

\bibitem{Sar2}
P.E. Sartwell, The incubation period and the dynamics of infectious disease,
\emph{American Journal of Epidemiology} \textbf{83} (1966) 204--216.

 \bibitem{SGD} E. Scalas, U. Garibaldi and S. Donadio, Statistical equilibrium in the simple exchange games I. Methods of solution and
application to the Bennati--Dragulescu--Yakovenko (BDY) game, \emph{Eur. Phys. J. B} \textbf{53}   (2006) 267--272.

\bibitem{SW}
K. Sznajd--Weron and J.  Sznajd, Opinion evolution in closed community,  \emph{Int. J. Mod. Phys. C} \textbf{11}
(2000)  1157--1165.

\bibitem{Toda}
A.A. Toda, A note on the size distribution of consumption: more double Pareto than lognormal. \emph{Macroeconomic Dynamics}, \textbf{21} (6) (2017) 1508--1518.

\bibitem{To3}
G. Toscani, Sur l'in\'egalit\'e logarithmique de Sobolev,  \emph{CRAS} \textbf{324}, S\'erie I (1997)  689--694.

\bibitem{To99}
G. Toscani, Entropy production and the rate of convergence to equilibrium for the Fokker-Planck equation, \emph{Quarterly of Appl. Math.} \textbf{LVII} (1999) 521--541.



\bibitem{To1}
G. Toscani, Kinetic models of opinion formation, \emph{Commun.
Math. Sci.} \textbf{4}  (2006) 481--496.


\bibitem{TBD} 
G. Toscani, C. Brugna and S. Demichelis, Kinetic models for the trading of goods, \emph{J. Stat. Phys}, \textbf{151} (2013)  549--566.

\bibitem{To2}
 G. Toscani, Kinetic and mean field description of Gibrat's law, \emph{ Physica A},  \textbf{461}  (2016)  802--811 

\bibitem{To4} 
G. Toscani,
Sharp weighted inequalities for probability densities on the real line, in preparation.

\bibitem{UM}
R. Ulrich and J. Miller, Information processing models generating
lognormally distributed reaction times, \emph{Journal of Mathematical
Psychology} \textbf{37} (1993) 513--525.

\bibitem{Vi}
C. Villani,
   Contribution {\`a} l'{\'e}tude math{\'e}matique des
  {\'e}quations de {B}oltzmann et de {L}andau en th{\'e}orie cin{\'e}tique des
  gaz et des plasmas.  {\em PhD thesis, Univ. Paris-Dauphine} (1998).

\bibitem{Zipf}
G.K. Zipf, \emph{Human behavior and the principle of least effort: An introduction
to human ecology} Addison-Wesley, Reading, MA 1949.




\end{thebibliography}
\end{document}